\documentclass[prx,aps,amssymb,twocolumn,nofootinbib,superscriptaddress]{revtex4-1}
\usepackage{amsmath}
\usepackage{amssymb}
\usepackage{amsthm}
\usepackage{amsfonts}
\usepackage{listings}
\usepackage[colorlinks, citecolor=blue]{hyperref}
\usepackage{dsfont}
\usepackage{enumerate}
\usepackage{latexsym}
\usepackage{longtable}
\usepackage[normalem]{ulem}
\usepackage{float}
\usepackage{multirow}

\usepackage{textgreek}

\usepackage{psfrag}

\usepackage{bm}
\usepackage{graphicx}
\usepackage{xcolor}

\newcommand{\beq}{\begin{equation}}
\newcommand{\eneq}{\end{equation}}

\input{epsf}

\def\beq#1\eeq{\begin{equation}#1\end{equation}}
\def\beqs#1\eeqs{\begin{align}#1\end{align}}

\newcommand{\bpm}{\begin{pmatrix}}
\newcommand{\epm}{\end{pmatrix}}

\newcommand{\bal}{\begin{align}}

\begin{document}

\newcommand{\TQCDBNbrICSDs}{73,234}
\newcommand{\TQCDBNbrNoSOCICSDs}{69,730}
\newcommand{\TQCDBPercentNoSOCICSDs}{95.22\%}
\newcommand{\TQCDBNbrFailedNoSOCICSDs}{3,504}
\newcommand{\TQCDBPercentFailedNoSOCICSDs}{4.78\%}
\newcommand{\TQCDBNbrUniqueMaterials}{38,298}
\newcommand{\TQCDBNbrNoSOCUniqueMaterials}{36,163}
\newcommand{\TQCDBPercentageNoSOCUniqueMaterials}{94.43\%}
\newcommand{\TQCDBNbrFailedNoSOCUniqueMaterials}{2,135}
\newcommand{\TQCDBPercentageFailedNoSOCUniqueMaterials}{5.57\%}

\newcommand{\TQCDBNbrMaterialsFElectrons}{10,987}
\newcommand{\TQCDBNbrMaterialsFElectronsPercent}{28.69\%}
\newcommand{\TQCDBNbrNoSOCICSDsNoFElectrons}{52,517}
\newcommand{\TQCDBNbrNoSOCICSDsNoFElectronsPercent}{75.31\%}
\newcommand{\TQCDBNbrMaterialsMagneticMP}{7,124}
\newcommand{\TQCDBNbrMaterialsMagneticMPPercent}{18.60\%}
\newcommand{\TQCDBNbrMaterialsMagneticMPVASP}{13,718}
\newcommand{\TQCDBNbrMaterialsMagneticMPVASPPercent}{35.82\%}
\newcommand{\TQCDBNbrMaterialsMagneticMPVASPFElectrons}{19,987}
\newcommand{\TQCDBNbrMaterialsMagneticMPVASPFElectronsPercent}{52.19\%}

\newcommand{\TQCDBNbrMaterialsTI}{6,128}
\newcommand{\TQCDBNbrMaterialsTIPercent}{16.00\%}
\newcommand{\TQCDBNbrMaterialsSM}{14,037}
\newcommand{\TQCDBNbrMaterialsSMPercent}{36.65\%}
\newcommand{\TQCDBNbrMaterialstrivial}{18,133}
\newcommand{\TQCDBNbrMaterialstrivialPercent}{47.35\%}

\newcommand{\TQCDBNbrMaterialsNLC}{3,000}
\newcommand{\TQCDBNbrMaterialsNLCPercent}{7.83\%}
\newcommand{\TQCDBNbrMaterialsSEBR}{3,128}
\newcommand{\TQCDBNbrMaterialsSEBRPercent}{8.17\%}
\newcommand{\TQCDBNbrMaterialsES}{4,102}
\newcommand{\TQCDBNbrMaterialsESPercent}{10.71\%}
\newcommand{\TQCDBNbrMaterialsESFD}{9,935}
\newcommand{\TQCDBNbrMaterialsESFDPercent}{25.94\%}
\newcommand{\TQCDBNbrMaterialsLCEBR}{18,133}
\newcommand{\TQCDBNbrMaterialsLCEBRPercent}{47.35\%}
\newcommand{\TQCDBNbrTopologicalMaterials}{20,165}
\newcommand{\TQCDBNbrTopologicalMaterialsPercent}{52.65\%}

\newcommand{\TQCDBNbrNoSOCMaterialsSM}{20,298}
\newcommand{\TQCDBNbrNoSOCMaterialsSMPercent}{56.13\%}
\newcommand{\TQCDBNbrNoSOCMaterialstrivial}{15,865}
\newcommand{\TQCDBNbrNoSOCMaterialstrivialPercent}{43.87\%}
\newcommand{\TQCDBNbrNoSOCMaterialsES}{6,006}
\newcommand{\TQCDBNbrNoSOCMaterialsESPercent}{16.61\%}
\newcommand{\TQCDBNbrNoSOCMaterialsESFD}{13,997}
\newcommand{\TQCDBNbrNoSOCMaterialsESFDPercent}{38.71\%}
\newcommand{\TQCDBNbrNoSOCMaterialsLCEBR}{15,865}
\newcommand{\TQCDBNbrNoSOCMaterialsLCEBRPercent}{43.87\%}
\newcommand{\TQCDBNbrNoSOCMaterialsNLCSM}{251}
\newcommand{\TQCDBNbrNoSOCMaterialsNLCSMPercent}{0.69\%}
\newcommand{\TQCDBNbrNoSOCMaterialsSEBRSM}{44}
\newcommand{\TQCDBNbrNoSOCMaterialsSEBRSMPercent}{0.12\%}
\newcommand{\TQCDBNbrNoSOCMaterialsNLCSMES}{6,257}
\newcommand{\TQCDBNbrNoSOCMaterialsSEBRSMESFD}{14,041}
\newcommand{\TQCDBNbrNoSOCMaterialsNLCSMSEBRSM}{295}

\newcommand{\TQCDBNbrMaterialsWithSOCNoSOCTI}{5,382}
\newcommand{\TQCDBNbrMaterialsWithSOCNoSOCTIPercent}{14.88\%}
\newcommand{\TQCDBNbrMaterialsWithSOCNoSOCSM}{13,270}
\newcommand{\TQCDBNbrMaterialsWithSOCNoSOCSMPercent}{36.69\%}
\newcommand{\TQCDBNbrMaterialsWithSOCNoSOCtrivial}{17,511}
\newcommand{\TQCDBNbrMaterialsWithSOCNoSOCtrivialPercent}{48.42\%}

\newcommand{\TQCDBNbrMaterialsWithSOCNoSOCNLC}{2,568}
\newcommand{\TQCDBNbrMaterialsWithSOCNoSOCNLCPercent}{7.10\%}
\newcommand{\TQCDBNbrMaterialsWithSOCNoSOCSEBR}{2,814}
\newcommand{\TQCDBNbrMaterialsWithSOCNoSOCSEBRPercent}{7.78\%}
\newcommand{\TQCDBNbrMaterialsWithSOCNoSOCES}{3,785}
\newcommand{\TQCDBNbrMaterialsWithSOCNoSOCESPercent}{10.47\%}
\newcommand{\TQCDBNbrMaterialsWithSOCNoSOCESFD}{9,485}
\newcommand{\TQCDBNbrMaterialsWithSOCNoSOCESFDPercent}{26.23\%}
\newcommand{\TQCDBNbrMaterialsWithSOCNoSOCLCEBR}{17,511}
\newcommand{\TQCDBNbrMaterialsWithSOCNoSOCLCEBRPercent}{48.42\%}
\newcommand{\TQCDBNbrMaterialsWithSOCNoSOCNLCSEBR}{5,382}
\newcommand{\TQCDBNbrMaterialsWithSOCNoSOCSEBRESFD}{12,299}
\newcommand{\TQCDBNbrMaterialsWithSOCNoSOCNLCES}{6,353}

\newcommand{\TQCDBNbrSTopo}{769}
\newcommand{\TQCDBNbrSTopoPercentage}{2.01\%}
\newcommand{\TQCDBNbrSTopoNLC}{37}
\newcommand{\TQCDBNbrSTopoNLCPercent}{0.10\%}
\newcommand{\TQCDBNbrSTopoSEBR}{109}
\newcommand{\TQCDBNbrSTopoSEBRPercent}{0.28\%}
\newcommand{\TQCDBNbrSTopoES}{178}
\newcommand{\TQCDBNbrSTopoESPercent}{0.46\%}
\newcommand{\TQCDBNbrSTopoESFD}{339}
\newcommand{\TQCDBNbrSTopoESFDPercent}{0.89\%}
\newcommand{\TQCDBNbrSTopoLCEBR}{106}
\newcommand{\TQCDBNbrSTopoLCEBRPercent}{0.28\%}

\newcommand{\TQCDBNbrTopoBand}{33,698}
\newcommand{\TQCDBNbrTopoBandPercent}{87.99\%}

\newcommand{\TQCDBNbrSMetal}{17}
\newcommand{\TQCDBNbrSMetalPercent}{0.04\%}
\newcommand{\TQCDBNbrSMetalICSDs}{65}
\newcommand{\TQCDBNbrSMetalPercentICSDs}{0.09\%}

\newcommand{\TQCDBBandSetsUniqueMaterials}{1,996,728}
\newcommand{\TQCDBNbrLCEBRBandSetsUniqueMaterials}{750,504}
\newcommand{\TQCDBPercentLCEBRBandSetsUniqueMaterials}{37.59\%}
\newcommand{\TQCDBNbrNLCBandSetsUniqueMaterials}{859,606}
\newcommand{\TQCDBPercentNLCBandSetsUniqueMaterials}{43.05\%}
\newcommand{\TQCDBNbrSEBRBandSetsUniqueMaterials}{379,321}
\newcommand{\TQCDBPercentSEBRBandSetsUniqueMaterials}{19.00\%}
\newcommand{\TQCDBNbrStrongBandSetsUniqueMaterials}{1,238,927}
\newcommand{\TQCDBPercentStrongBandSetsUniqueMaterials}{62.05\%}
\newcommand{\TQCDBNbrFragileBandSetsUniqueMaterials}{7,297}
\newcommand{\TQCDBPercentFragileBandSetsUniqueMaterials}{0.37\%}

\newcommand{\TQCDBNbrNoSOCSTopo}{28}
\newcommand{\TQCDBNbrNoSOCSTopoPercentage}{0.08\%}
\newcommand{\TQCDBNbrNoSOCSTopoNLC}{0}
\newcommand{\TQCDBNbrNoSOCSTopoNLCPercent}{0.00\%}
\newcommand{\TQCDBNbrNoSOCSTopoSEBR}{0}
\newcommand{\TQCDBNbrNoSOCSTopoSEBRPercent}{0.00\%}
\newcommand{\TQCDBNbrNoSOCSTopoES}{5}
\newcommand{\TQCDBNbrNoSOCSTopoESPercent}{0.01\%}
\newcommand{\TQCDBNbrNoSOCSTopoESFD}{14}
\newcommand{\TQCDBNbrNoSOCSTopoESFDPercent}{0.04\%}
\newcommand{\TQCDBNbrNoSOCSTopoLCEBR}{8}
\newcommand{\TQCDBNbrNoSOCSTopoLCEBRPercent}{0.02\%}

\newcommand{\TQCDBNbrNoSOCTopoBand}{10,001}
\newcommand{\TQCDBNbrNoSOCTopoBandPercent}{27.66\%}

\newcommand{\TQCDBNbrNoSOCSMetal}{1,138}
\newcommand{\TQCDBNbrNoSOCSMetalPercent}{3.15\%}
\newcommand{\TQCDBNbrNoSOCSMetalICSDs}{3,495}
\newcommand{\TQCDBNbrNoSOCSMetalPercentICSDs}{5.01\%}

\newcommand{\TQCDBVASPICPUhSOC}{9,500,801.70}
\newcommand{\TQCDBVASPIICPUhSOC}{170,633.80}
\newcommand{\TQCDBVASPIIICPUhSOC}{2,560,882.60}
\newcommand{\TQCDBVASPIVCPUhSOC}{5,804,948.80}
\newcommand{\TQCDBVASPTotalCPUhSOC}{18,037,266.90}

\newcommand{\TQCDBVASPICPUhNoSOC}{2,298,583.30}
\newcommand{\TQCDBVASPIICPUhNoSOC}{56,520.70}
\newcommand{\TQCDBVASPIIICPUhNoSOC}{773,019.50}
\newcommand{\TQCDBVASPIVCPUhNoSOC}{1,397,550.30}
\newcommand{\TQCDBVASPTotalCPUhNoSOC}{4,525,673.80}

\newcommand{\TQCDBVASPTotalCPUTimeAllHours}{22,562,940.70}
\newcommand{\TQCDBVASPTotalCPUTimeAllMillionHours}{22.60}

\newcommand{\TQCDBTotalStorage}{2,037.80Gb}

\newcommand{\TQCDBNoSOCTotalStorage}{343.30Gb}
\newcommand{\TQCDBPercentNoSOCTotalStorage}{16.85\%}
\newcommand{\TQCDBNoSOCPROCARStorage}{173.60Gb}
\newcommand{\TQCDBNoSOCCHGCARStorage}{119.00Gb}

\newcommand{\TQCDBTotalSOCStorage}{1,694.50Gb}
\newcommand{\TQCDBPercentSOCTotalStorage}{83.15\%}
\newcommand{\TQCDBPROCARSOCStorage}{1,013.80Gb}
\newcommand{\TQCDBPercentPROCARSOCStorage}{49.75\%}
\newcommand{\TQCDBCHGCARSOCStorage}{559.30Gb}
\newcommand{\TQCDBPercentCHGCARSOCStorage}{27.45\%}

\newcommand{\TQCDBNbrSkippedICSDs}{22,996}
\newcommand{\TQCDBNbrFailedComputedICSDs}{11,315}
\newcommand{\TQCDBNbrFailedVASPToTraceICSDs}{5,014}
\newcommand{\TQCDBNbrFailedVASPToTraceBadTRICSDs}{1,703}
\newcommand{\TQCDBNbrFailedVASPToTraceBadTracesICSDs}{655}
\newcommand{\TQCDBNbrFailedVASPToTraceAccidentalFermiICSDs}{2,647}


\newcommand{\icsdwebshort}[1]{\href{https://www.topologicalquantumchemistry.com/\#/detail/#1}{#1}}
\newcommand{\icsdweb}[1]{\href{https://www.topologicalquantumchemistry.com/\#/detail/#1}{ICSD #1}}
\newcommand{\webNoICSD}{\url{https://www.topologicalquantumchemistry.com/}}
\newcommand{\webTQC}{\href{https://www.topologicalquantumchemistry.com/}{Topological Materials Database}}

\newcommand{\webBCSfull}{\href{https://www.cryst.ehu.es/}{Bilbao Crystallographic Server}}

\newcommand{\webBCSshort}{\href{https://www.cryst.ehu.es/}{BCS}}

\newcommand{\webchecktopmat}{\href{https://www.cryst.ehu.es/cryst/checktopologicalmat}{Check Topological Mat}}
\newcommand{\identify}{\href{www.cryst.ehu.es/cryst/identify_group}{IDENTIFY GROUP}}

\newcommand{\vasptotrace}{\href{https://github.com/zjwang11/irvsp}{VASP2Trace}}
\newcommand{\checktopmat}{\href{https://www.cryst.ehu.es/cryst/checktopologicalmat}{Check Topological Mat}}

\newcommand{\TQCDTotICSDs}{193,426}
\newcommand{\TQCDTotICSDsExp}{181,218}
\newcommand{\TQCDTotICSDsTheo}{12,208}
\newcommand{\ThresholdNbrAtoms}{60}
\newcommand{\TQCDstoichiometric}{96,196}
\newcommand{\TQCDstoichiometricPercent}{49.73\%}
\newcommand{\TQCDstoichiometricExp}{85,701}
\newcommand{\TQCDstoichiometricExpPercent}{89.09\%}
\newcommand{\TQCDstoichiometricTheo}{10,495}
\newcommand{\TQCDstoichiometricTheoPercent}{10.91\%}
\newcommand{\TQCDNostoichiometric}{97,230}
\newcommand{\TQCDNostoichiometricPercent}{50.27\%}
\newcommand{\TQCDNostoichiometricExp}{95,517}
\newcommand{\TQCDNostoichiometricExpPercent}{98.24\%}
\newcommand{\TQCDNostoichiometricTheo}{1,713}
\newcommand{\TQCDNostoichiometricTheoPercent}{1.76\%}
\newcommand{\TQCDstoichiometricLEQatoms}{85,361}
\newcommand{\TQCDstoichiometricLEQatomsPercent}{88.74\%}
\newcommand{\TQCDstoichiometricExpLEQatoms}{75,375}
\newcommand{\TQCDstoichiometricExpLEQatomsPercent}{87.95\%}
\newcommand{\TQCDstoichiometricTheoLEQatoms}{9,986}
\newcommand{\TQCDstoichiometricTheoLEQatomsPercent}{95.15\%}
\newcommand{\TQCDstoichiometricGTatoms}{10,835}
\newcommand{\TQCDstoichiometricGTatomsPercent}{11.26\%}
\newcommand{\TQCDstoichiometricExpGTatoms}{10,326}
\newcommand{\TQCDstoichiometricExpGTatomsPercent}{12.05\%}
\newcommand{\TQCDstoichiometricTheoGTatoms}{509}
\newcommand{\TQCDstoichiometricTheoGTatomsPercent}{4.85\%}

\newcommand{\TQCDBHighConnectivityNbrBandsThreshold}{130}
\newcommand{\TQCDBNbrHighConnectivityMaterials}{244}
\newcommand{\TQCDBNbrNoSOCHighConnectivityMaterials}{972}
\newcommand{\TQCDBNbrMaterialsBndLowerEf}{508}
\newcommand{\TQCDBNbrMaterialsBndLowerEfICSDs}{1,047}
\newcommand{\TQCDBNbrNoSOCMaterialsBndLowerEf}{10,313}
\newcommand{\TQCDBNbrNoSOCMaterialsBndLowerEfICSDs}{20,899}

\newcommand{\vaspcputimelabel}[1]{\begin{tabular}{c}VASP#1\\(CPU hours)\end{tabular}}

\newcommand{\supappref}[1]{SA~\ref{#1}}

\newcommand{\webmaterialsproject}{\href{https://materialsproject.org}{Materials Project}}

\title{Topological Materials Discovery from Crystal Symmetry}

\author{Benjamin J. Wieder$^\dag$}
\affiliation{Department of Physics, Massachusetts Institute of Technology, Cambridge, MA 02139, USA}
\affiliation{Department of Physics, Northeastern University, Boston, MA 02115, USA}
\affiliation{Department of Physics,
Princeton University,
Princeton, NJ 08544, USA
	}

\author{Barry Bradlyn}
\affiliation{Department of Physics and Institute for Condensed Matter Theory, University of Illinois at Urbana-Champaign, Urbana, IL, 61801-3080, USA}

\author{Jennifer Cano}
\affiliation{Department of Physics and Astronomy, Stony Brook University, Stony Brook, New York 11974, USA}
\affiliation{Center for Computational Quantum Physics, The Flatiron Institute, New York, New York 10010, USA}

\author{Zhijun Wang}
\affiliation{Beijing National Laboratory for Condensed Matter Physics, and Institute of Physics, Chinese Academy of Sciences, Beijing 100190, China}
\affiliation{University of Chinese Academy of Sciences, Beijing 100049, China}

\author{Maia G. Vergniory}
\affiliation{Donostia International Physics Center, P. Manuel de Lardizabal 4, 20018 Donostia-San Sebastian, Spain}
\affiliation{IKERBASQUE, Basque Foundation for Science, Bilbao, Spain}
\affiliation{Max Planck Institute for Chemical Physics of Solids, D-01187 Dresden, Germany}

\author{Luis Elcoro}
\affiliation{Department of Condensed Matter Physics, 
University of the Basque Country UPV/EHU, 
Apartado 644, 48080 Bilbao, Spain}

\author{Alexey A. Soluyanov}
\thanks{Deceased 26 October 2019}
\affiliation{Department of Physics, University of Zurich, Winterthurerstrasse 190, 8057 Zurich, Switzerland}
\affiliation{Department of Physics, St. Petersburg State University, St. Petersburg, 199034, Russia}

\author{Claudia Felser}
\affiliation{Max Planck Institute for Chemical Physics of Solids, D-01187 Dresden, Germany}

\author{Titus Neupert}
\affiliation{Department of Physics, University of Zurich, Winterthurerstrasse 190, 8057 Zurich, Switzerland}

\author{Nicolas Regnault}
\affiliation{Laboratoire de Physique de l'{\'E}cole Normale Sup{\'e}rieure, PSL University, CNRS, Sorbonne Universit{\'e}, Universit{\'e} Paris Diderot, Sorbonne Paris Citv{\'e}, Paris, France}
\affiliation{Department of Physics,
Princeton University,
Princeton, NJ 08544, USA
	}

\author{B. Andrei Bernevig$^\dag$$^\ddag$}
\affiliation{Department of Physics,
Princeton University,
Princeton, NJ 08544, USA
	}
\affiliation{Donostia International Physics Center, P. Manuel de Lardizabal 4, 20018 Donostia-San Sebastian, Spain}
\affiliation{IKERBASQUE, Basque Foundation for Science, Bilbao, Spain}

\date{\today}

\begin{abstract}
Topological materials discovery has evolved at a rapid pace over the past 15 years following the identification of the first nonmagnetic topological insulators (TIs), topological crystalline insulators (TCIs), and 3D topological semimetals (TSMs).  Most recently, through complete analyses of symmetry-allowed band structures -- including the theory of Topological Quantum Chemistry (TQC) -- researchers have determined crystal-symmetry-enhanced Wilson-loop and complete symmetry-based indicators for nonmagnetic topological phases, leading to the discovery of higher-order TCIs and TSMs.  The recent application of TQC and related methods to high-throughput materials discovery has revealed that over half of all of the known stoichiometric, solid-state, nonmagnetic materials are topological at the Fermi level, over 85\% of the known stoichiometric materials host energetically isolated topological bands, and that just under $2/3$ of the energetically isolated bands in known materials carry the stable topology of a TI or TCI.  In this Review, we survey topological electronic materials discovery in nonmagnetic crystalline solids from the prediction of the first 2D and 3D TIs to the recently introduced methods that have facilitated large-scale searches for topological materials.  We also discuss future venues for the identification and manipulation of solid-state topological phases, including charge-density-wave compounds, magnetic materials, and 2D few-layer devices.
\end{abstract}

\maketitle


\section{Introduction}

Nearly 15 years have passed since the first theoretical predictions of the quantum spin Hall effect~\cite{CharlieTI,KaneMeleZ2,AndreiTI} in what are now known as 2D topological insulators (TIs).  TIs, as well as topological semimetals (TSMs), are unique solid-state systems that exhibit robust boundary states and quantized bulk response coefficients as a consequence of their electronic wavefunctions.  Because topological boundary states survive under moderate disorder and frequently exhibit spin polarization and spin-momentum locking, topological materials represent a promising possible venue for storing and manipulating quantum information~\cite{Kitaev1,Kitaev2,FuKaneSC}, and realizing coherent spin transport~\cite{TISpintronic1,TISpintronic2,TISpintronicReview} and high-efficiency catalysis~\cite{ClaudiaWeylCatalysis,ClaudiaChiralCatalysis,OtherChiralCatalysis}.  Since the discovery of the first TIs, a large number of both insulating and (semi)metallic topological phases have been predicted, characterized, and measured in experimentally accessible chemical compounds.  This unprecedented pace of topological materials discovery has been driven by several major factors.  From a theoretical perspective, the discovery of novel topological phases has been facilitated by the rapid development of topological invariants computable from band structures, and from the rediscovery and application of crystalline symmetry and group theory.  The growth of topological condensed matter physics can also be attributed to the major advances in cloud storage and large-scale computing that have enabled high-throughput calculations, and to dramatic improvements in network bandwidth and digital communication that have enabled close collaboration between international teams of theorists and experimentalists.

In this work, we review topological materials discovery from the identification of the first TIs and TSMs to the modern, ``Renaissance'' era of high-throughput topological classification and materials databases.  For each stoichiometric 3D material discussed in this Review, we have provided the~\href{https://icsd.products.fiz-karlsruhe.de/}{Inorganic Crystal Structure Database (ICSD)}~\cite{ICSD1,ICSD2} accession code of the material with a hyperlink to the associated page on the~\webTQC~(\webNoICSD)~\cite{AndreiMaterials,AndreiMaterials2}, as well as the number and standard-setting symbol of the crystallographic space group (SG) of the material.

Owing to the sheer volume of impactful discoveries made since the prediction of 2D TIs, some topological materials and classifications are only lightly covered or omitted from this Review.  Interested readers may further consult a number of previous reviews~\cite{CharlieReview,MarzariReview,WeylReview,RappeReview,BinghaiReview,NagaosaWeylReview}.  Additionally, in this Review, we focus on the discovery and characterization of topology in the electronic structure of crystalline chemical compounds with weak interactions and negligible disorder.  As such, we do not discuss the topological characterization of metamaterials, sonic and Floquet crystals, superconductors, classical mechanical systems, magnonic bands, and strongly disordered systems, which have also seen tremendous advances in recent years.

\section{Topological Band Theory}

In this section, we first introduce the band theory of crystalline solids, and then detail methods through which an energetically isolated set of bands can be diagnosed as topologically nontrivial.  Finally, we address experimentally observable signatures of topological bands in solid-state materials.

\subsection{The Band Theory of Crystalline Solids}
\label{sec:bands}

From a theoretical perspective, a crystal is defined as a system in which physical observables, such as the energy spectrum and the locations of atoms, are periodic under the action of discrete translations.  From an experimental perspective, the picture is more complicated: samples of crystalline materials carry defects, misoriented domains, and coexisting structural orders, such as left-handed and right-handed enantiomers~\cite{AshcroftMermin,KaustuvReview,MerminReview,FlackChirality}.  In this subsection, we focus on the underlying mathematical machinery used to characterize the electronic states in theoretical models of pristine, solid-state crystals.  In Sec.~\ref{sec:introTopology}, we then briefly generalize to models that incorporate sources of deviation from the perfect-crystal limit, such as disorder.

\begin{figure}[!t]
\centering
\includegraphics[width=\columnwidth]{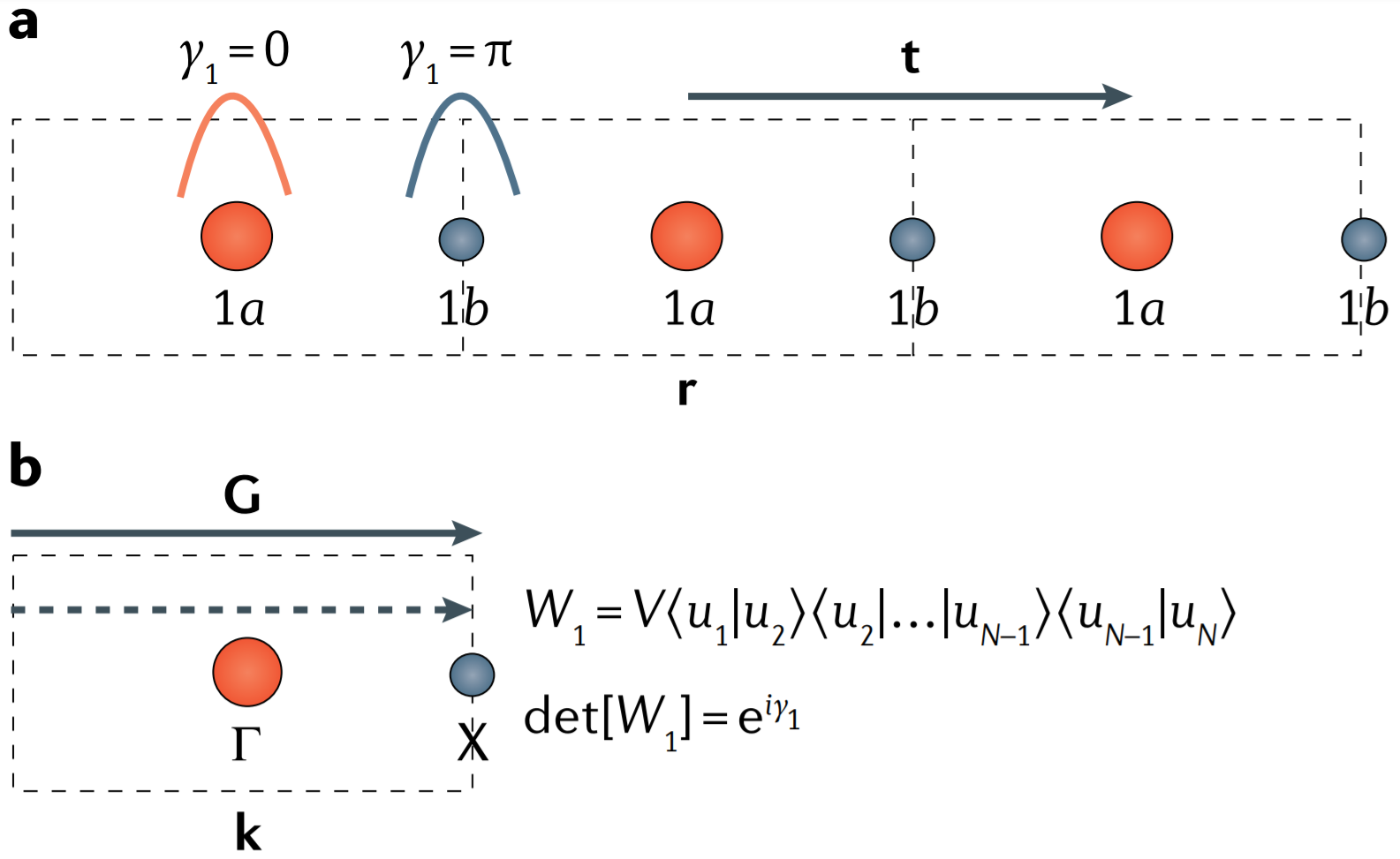}
\caption{Wannier centers and Wilson loops.  (a) A 1D crystal in position (${\bf r}$) space, where adjacent unit cells (black dashed rectangles) are related through lattice translation by the vector ${\bf t}$.  The orange and gray curves at the $1a$ and $1b$ Wyckoff positions in the left-most unit cell symbolize exponentially localized, inversion- ($\mathcal{I}$-) symmetric Wannier orbitals~\cite{MarzariReview,MarzariDavidWannier,RestaReview} obtained from inverse-Fourier-transforming bands with a Berry phase~\cite{BerryPhase} (equivalently termed a Zak phase~\cite{ZakPhase}) $\gamma_{1}=0$ and $\gamma_{1}=\pi$, respectively.  (b) ${\bf k}$ points in the first Brillouin zone (BZ) of the 1D crystal in (a) are related to ${\bf k}$ points in the second BZ by the reciprocal lattice vector ${\bf G}$.  For a set of occupied bands $|u_{\bf k}\rangle$ at each point ${\bf k}$ in the first BZ, the discretized Wilson loop matrix $W_{1}$ is defined as the product of the outer products of $|u_{\bf k}\rangle$ and $|u_{\bf k+1}\rangle$ taken in the direction of the dashed arrow over a finite mesh of ${\bf k}$ points in the first BZ (as well as a sewing matrix $V$)~\cite{Fidkowski2011,AndreiXiZ2,AlexeyWilson,ArisInversion}.  Crucially, for an energetically isolated set of bands, the total Berry phase $\gamma_{1}\text{ mod }2\pi$ is related to the Wilson loop matrix by $\det[W_{1}]=\exp(i\gamma_{1})$.}
\label{fig:Wilson}
\end{figure}

To begin, in a theoretical model of a pristine, solid-state crystal, electronic phenomena are captured through a position- (${\bf r}$-) space Hamiltonian that is periodic with respect to the group of 3D lattice translations~\cite{BigBook} [Fig.~\ref{fig:Wilson}(a)].  In this Review, we focus on noninteracting materials, whose energy spectra and electronic response properties are largely captured in the mean-field approximation in which electron-electron correlations are either incorporated as background fields or taken to vanish.  Whereas the noninteracting electronic Hamiltonian $\mathcal{H}({\bf r})$ generically contains off-diagonal elements that couple states at different positions ${\bf r}$ and ${\bf r}'$, the Fourier-transformed Bloch Hamiltonian $\mathcal{H}({\bf k})$ is diagonal in ${\bf k}$.  The energy eigenvalues of the Bloch Hamiltonian $\mathcal{H}({\bf k})$ form bands that are periodic with respect to advancing the crystal momentum ${\bf k}$ by a reciprocal lattice vector ${\bf G}$: ${\bf k}\rightarrow {\bf k}+{\bf G}$ [Fig.~\ref{fig:Wilson}(b)].  In ${\bf k}$-space, the unit cells of the reciprocal lattice are known as the Brillouin zones (BZs)~\cite{AshcroftMermin,BigBook}.  In solid-state materials, the bands are populated by electrons up to the Fermi energy ($E_{F}$).  Hence, crystals are filled with spinful fermions (electrons), even in the limit of negligible coupling between the atomic orbitals and electron spins [spin-orbit coupling (SOC)].

In addition to translation symmetry, 3D crystals may respect other unitary symmetries, such as spatial inversion ($\mathcal{I}$), or antiunitary symmetries, such as time-reversal ($\mathcal{T}$).  The set of symmetries that leave the orbitals (and electron spins) in an infinite crystal invariant (up to integer lattice translations) form a group; whether it is a space, layer, wallpaper, rod, or other group depends on the dimensionalities of lattice translations and crystal symmetries~\cite{ConwayThings}.  The symmetries of nonmagnetic 3D crystals are specifically contained within the 230 nonmagnetic ($\mathcal{T}$-symmetric) SGs~\cite{BigBook}.  Crucially, unitary particle-hole (also known as chiral) symmetry, which relates the valence and conduction bands, is generically absent in real materials.  The accidental or intentional incorporation of chiral symmetry into topological models is a common source of spurious predictions.  Therefore, chiral symmetry, while a helpful bootstrap for forming simple models, should be carefully relaxed when predicting robust topological phenomena.

At each ${\bf k}$ point, bands can be uniquely labeled by the simultaneous eigenvalues (quantum numbers) of commuting unitary symmetry operations.  The independent combinations of simultaneous symmetry eigenvalues can then be inferred from the symmetry operation matrix representatives in the small irreducible representations of the symmetry (little) group of each ${\bf k}$ point~\cite{BigBook}.  When the actions of antiunitary symmetries are incorporated, irreducible representations may become paired into irreducible corepresentations (coreps)~\cite{Wigner1932,wigner1959group}.  In crystals with antiunitary symmetries -- most notably including nonmagnetic solid-state materials -- the bands at each ${\bf k}$ point are labeled by small coreps; the small coreps corresponding to all possible electronic states in nonmagnetic crystals were tabulated in~\cite{BigBook}, and reproduced in the~\href{http://www.cryst.ehu.es/cgi-bin/cryst/programs/representations.pl?tipogrupo=dbg}{REPRESENTATIONS DSG} tool on the~\webBCSfull~(\webBCSshort)~\cite{QuantumChemistry,Bandrep1,BCS1,BCS2}.

For a group of bands that is separated from all of the other bands in the energy spectrum by an energy gap at each ${\bf k}$ point, one may attempt to inverse-Fourier transform the bands into exponentially localized Wannier orbitals in position space [Fig~\ref{fig:Wilson}(a)].  If a group of bands can be inverse-Fourier transformed into exponentially localized Wannier orbitals that respect the symmetries of the crystal SG, then the bands are termed Wannierizable~\cite{MarzariReview,MarzariDavidWannier,RestaReview}.  As we discuss in the next subsection, the absence of Wannierizability, or the properties of Wannier functions if successfully computed, are central to the identification of nontrivial band topology.

\subsection{Diagnosing Band Topology}
\label{sec:introTopology}

From a mathematical perspective, the properties of an object are considered to be topological if they do not change as the object is smoothly deformed.  In some cases, symmetry-protected topological properties may also be defined by restricting consideration to objects and deformations that respect a given set of symmetries.  Just under 40 years ago~\cite{TKNN,ThoulessPump,ThoulessWannier}, it was discovered that, remarkably, the Bloch wavefunctions of energetically isolated bands in crystalline solids can carry topologically quantized numbers -- known as topological invariants -- that correspond to experimentally measurable quantized response effects~\cite{VonKlitzingNobel}.  The computation of topological invariants has allowed researchers to predict identically quantized responses in seemingly disparate solid-state materials, as well as in highly-simplified theoretical models.  This is perhaps unintuitive, because the dispersion of energetically isolated bands in a material is free to change under doping, strain, or in the presence of weak disorder.  Hence, one might expect little correspondence between the generic properties of simple models and those of real chemical compounds.  However, if the band distortion and interactions do not close a gap or break a symmetry, then the topology of the isolated bands cannot change.  Because topology and symmetry alone indicate the numbers and locations of boundary states and the quantized values of bulk response coefficients in many solid-state materials, topological phenomena can be meaningfully predicted using appropriately simplified models.

Originally, nontrivial band topology was considered in the context of bulk-insulating systems~\cite{TKNN,ThoulessPump,ThoulessWannier}.  However, the only requirement for unambiguously diagnosing the topology of a group of bands is that the bulk bands be energetically separated from the other bands in the spectrum at each momentum, such that there is a gap at each ${\bf k}$ point; a direct gap or an insulating, indirect gap at a particular $E_{F}$ is not a prerequisite for nontrivial band topology.  Hence, as first emphasized by Liang Fu and Charles Kane in establishing the nontrivial band topology of metallic antimony [\icsdweb{426972}, SG 166 ($R\bar{3}m$)]~\cite{FuKaneInversion}, it is also possible to classify metallic materials as topological band insulators.  In this Review, we specifically define a band insulator as a material with $N_{e}$ electrons per unit cell in which the $N_{e}$ lowest-energy Bloch eigenstates at each ${\bf k}$ point are separated from all higher-energy Bloch eigenstates at ${\bf k}$ by a gap.  Therefore, we employ the convention established in~\cite{FuKaneInversion}, in which a material is classified as a TI or a TCI if the lowest $N_{e}$ bands in energy are as a set topological, even if there are bulk electron or hole Fermi pockets.

Topological material predictions are reliable because topological robustness is an extensive property: the larger the size of the topological bulk of a system in real space, the more robust the bulk topology is to interactions and disorder that preserve the system symmetry on the average.  This property crucially allows topological response effects and boundary states in simplified models of pristine crystals to be generalized to real-material samples, which inevitably host inhomogeneities such as crystal defects, dopants, and reconstructed surface textures~\cite{AshcroftMermin,KaustuvReview,MerminReview}.  For Chern insulators, extensive topological robustness has been demonstrated using network models~\cite{ChalkerCoddington,DungHaiNetwork,ChalkerHo} and renormalization group flow~\cite{PruiskenScaling}.  More recently, similar calculations were also performed to extract the critical conductance and correlation lengths of the Anderson localization transition in TIs and TCIs~\cite{TeoKanePinchOff,FuKaneFugacity,JensSurfaceWTI,RahulMBLWTI,AXIDelocalizeZhida,AXIDelocalizeXie}.

\vspace{0.1in}
\subsubsection{Classes of Topological Bands}
\label{sec:subsubBandClass}

Heuristically, there are three broad classes of topological bands in solid-state materials: obstructed atomic limits, fragile bands, and stable topological bands.  First, in obstructed atomic limits, the isolated bands are Wannierizable, but the Wannier centers reside at positions in the unit cell away from the underlying atomic ions~\cite{QuantumChemistry,JenOAL}.  Consequently, obstructed atomic limits exhibit electric multipole moments~\cite{VDBpolarization,RestaReview}, and boundaries between atomic limits and obstructed atomic limits exhibit either 0D or flat-band-like topological solitons with fractional charge or spin-charge separation~\cite{HeegerReview}.  In 1D, Wannier functions are always exponentially localized~\cite{KohnWannier} in the absence of superconductivity~\cite{FrankBarrySC}, and hence, all topological phases are obstructed atomic limits.  However, in 2D and 3D, one must be more careful in constructing Wannier functions, as topological bands in 2D and 3D may admit a localized ${\bf r}$-space description with power-law decaying tails, as opposed to the exponential decay required for Wannierizability~\cite{AlexeyVDBWannier,BarryFragile,WiederAxion}.  Topologically quantized charge and spin on domain walls between atomic limits were first identified in the Su-Schrieffer-Heeger~\cite{SSH,SSHspinon} and Rice-Mele~\cite{RiceMele} models of polyacetylene.  In the Su-Schrieffer-Heeger model, there are two insulating regimes whose occupied bands differ by a $\pi$ Berry~\cite{BerryPhase} phase -- equivalently termed a Zak~\cite{ZakPhase} phase -- that indicates the relative positions of the occupied Wannier orbitals~\cite{VDBpolarization,RestaReview} [Fig.~\ref{fig:Wilson}(a)].

\begin{figure*}[!t]
\centering
\includegraphics[width=\textwidth]{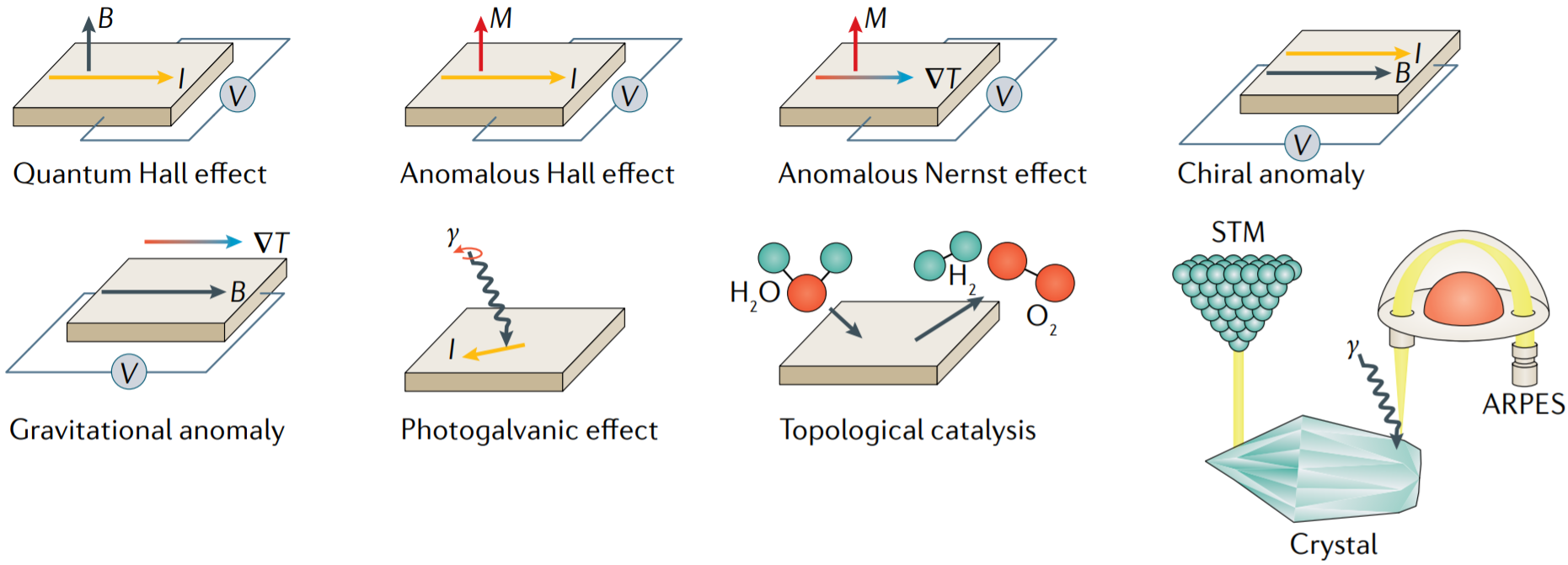}
\caption{Experimentally verified topological response effects in topological materials.  In the response effect schematics, $B$ is an external magnetic field, $M$ is magnetization, $I$ is electric current, $V$ is voltage, $T$ is temperature, and \textgamma\ is a photon.  Notable topological response effects include the quantum Hall effect (quantized $V$ induced by perpendicular $I$ and $B$)~\cite{VonKlitzingNobel}, the anomalous Hall effect ($V$ induced by perpendicular $I$ and $M$)~\cite{HaldaneModel,QAHScience1,QAHScience2}, the anomalous Nernst effect ($V$ induced by perpendicular $\nabla T$ and $M$)~\cite{ClaudiaNernst,NatPhysNernst1,NatPhysNernst2,PRMNernst,KaustuvReview}, the chiral anomaly ($I$ induced by parallel $V$ and $B$)~\cite{NielsenNinomiyaABJ,SonSpivakChiral,PhuonChiralDirac,ChiralAnomalyWeyl1,ChiralAnomalyWeyl2,ChiralAnomalyWeyl3,ZahidChiralAnomaly,DiracChiralAnomaly,JenChiralAnomaly}, the gravitational anomaly ($V$ induced by parallel $\nabla T$ and $B$)~\cite{ClaudiaGravitational,FranzReview}, the photogalvanic effect ($I$ induced by incident light)~\cite{WeylPhotogalvanic2017,AdolphoCPGE,FlickerMultifold,PennChiralCPGE1,SuyangPabloCPGEWTe2,CPGEMooreExp,PennChiralCPGE2}, and topologically enhanced catalysis (chemical reaction efficiency enhancement from topological surface states)~\cite{ClaudiaWeylCatalysis,ClaudiaChiralCatalysis,OtherChiralCatalysis}.  In 3D solid-state materials, experimental signatures of topological response effects are frequently bolstered by the concurrent observation of topological surface states in angle-resolved photoemission spectroscopy (ARPES)~\cite{ARPESReviewYulin,ARPESReviewHongDing} and scanning-tunneling microscopy (STM)~\cite{YazdaniQPI3DTI,AndreiAliMajoranaSTM,AliWeylQPI,MagneticWeylHaim,DemlerQPIMethod} experiments.}
\label{fig:response}
\end{figure*}

More recently, the polarization topology of obstructed atomic limits was extended to quadrupole and octupole moments through a formulation of nested Berry phase~\cite{multipole,WladTheory,HOTIBernevig,HingeSM,WiederAxion,TMDHOTI,WladCorners}.  First, for a set of occupied bands $|u_{\bf k}\rangle$ at each point ${\bf k}$ in the first BZ, one may define a discretized Wilson loop matrix $W_{1}$ as the product of the outer products of $|u_{\bf k}\rangle$ and $|u_{\bf k+1}\rangle$ taken over a finite mesh of ${\bf k}$ points in the first BZ (as well as a sewing matrix $V$)~\cite{Fidkowski2011,AndreiXiZ2,AlexeyWilson,ArisInversion}.  The total Berry phase $\gamma_{1}\text{ mod }2\pi$ of the occupied bands is then related to the Wilson loop matrix by $\det[W_{1}]=\exp(i\gamma_{1})$ [Fig.~\ref{fig:Wilson}(b)].  In this sense, the Wilson loop $W_{1}$ represents the non-Abelian (many-band) generalization of the Berry phase~\cite{Fidkowski2011,TRPolarization,AlexeyVDBWannier,AndreiXiZ2,AlexeyWilson,ArisInversion,BarryFragile,Cohomological,HourglassInsulator,DiracInsulator}.  It was recently recognized that because the eigenvalues of $W_{1}$ form smooth and continuous bands, then a second, nested Wilson loop matrix $W_{2}$ may be constructed by computing a perpendicular Wilson loop using the eigenstates of $W_{1}$.  Analogous to the relationship between the Berry phase $\gamma_{1}$ and $W_{1}$, a nested Berry phase $\gamma_{2}$ can then be defined through the relation $\det[W_{2}]=\exp(i\gamma_{2})$~\cite{multipole,WladTheory,HOTIBernevig,HingeSM,WiederAxion,TMDHOTI,WladCorners}.  The nested Berry phase $\gamma_{2}$ was first introduced to diagnose the topology of quantized quadrupole insulators, in which $\gamma_{2}=0,\pi$ indicates whether a pair of Wannier orbitals occupy the corner or the center of a 2D square lattice~\cite{multipole,WladTheory,HingeSM,ZackPhotonQI}.  Additionally, 3D obstructed atomic limits have recently been predicted in silicon crystals [\icsdweb{150530}, SG 227 ($Fd\bar{3}m$)]~\cite{ChamonSilicon} and in electrides including Y$_2$C [\icsdweb{672345}, SG 166 ($R\bar{3}m$)]~\cite{MurakamiElectride}, Li$_{12}$Al$_{3}$Si$_{4}$ [\icsdweb{39597}, SG 220 ($I\bar{4}3d$)]~\cite{AndreiElectride}, and Ca$_2$As [\icsdweb{166865}, SG 139 ($I4/mmm$)]~\cite{ElectrideNodalLine,ZhijunTQCElectride}.  The theory of topological quantum chemistry (TQC)~\cite{QuantumChemistry,JenFragile1,Bandrep1,Bandrep2,Bandrep3,BarryFragile}, which we discuss in detail in Sec.~\ref{sec:TQC}, is built upon the complete enumeration of atomic limits in the 230 nonmagnetic SGs.  In this Review, we primarily discuss the boundary states of obstructed atomic limits in the context of the flat-band-like surface and hinge states of TSMs.

It was recently discovered~\cite{AshvinFragile,JenFragile1,BarryFragile,AdrianFragile,ZhidaFragile,KoreanFragileInversion} that, in some cases, a group of bands exhibits boundary solitons~\cite{AshvinFragile2,HingeSM,TMDHOTI,WiederAxion,KoreanFragile,FragileFlowMeta,ZhidaFragile2}, but does not admit a Wannier description unless combined with other, trivial (Wannierizable) bands.  These fragile topological bands have been shown to lie close to $E_{F}$ in hundreds of solid-state materials~\cite{AndreiMaterials2}, and to be relevant to the low-energy excitations in twisted bilayer graphene~\cite{ZhidaBLG,AshvinBLG1,AshvinBLG2}.

The last class of bands -- and the most ubiquitous~\cite{AndreiMaterials2} -- exhibit the stable topology of TIs and TCIs.  Stable topological bands, unlike fragile bands or obstructed atomic limits, cannot be Wannierized, and remain non-Wannierizable when combined with (obstructed) atomic limits or fragile bands.  However, stable topological bands can generically be partially Fourier-transformed into hybrid Wannier functions with one coordinate $r_{\perp}$ in position space perpendicular to two components ${\bf k}_{\parallel}$ in momentum space~\cite{MarzariReview,MarzariDavidWannier,VDBSheet,Z2pack,NicoDavidAXI2}.  As first discovered by David Thouless, Mahito Kohomoto, Peter Nightingale, and Marcel de Nijs~\cite{TKNN,ThoulessPump,ThoulessWannier} in the 1980's, in stable TIs and TCIs, the positions of the hybrid Wannier functions necessarily wind as functions of ${\bf k}_{\parallel}$.  In a 2D quantum Hall phase considered in the limit of a rational flux quantum per unit cell such that there exists a 2D Bravais lattice of commuting (magnetic) translations~\cite{ZakMagneticTranslation}, Thouless, Kohomoto, Nightingale, de Nijs, and other researchers~\cite{NiuPump} recognized that the hybrid-Wannier-center winding can be classified by an integer-valued topological invariant termed a Chern number.  Crucially, the Chern number of a set of bands cannot be changed without closing a gap between the bands and other bands in the spectrum.  Additionally, the Chern number of all of the occupied bands corresponds to the bulk Hall conductivity $\sigma_{H}^{xy}$ in the units of $e^{2}/h$, as well as to the number of topological chiral edge states.

After the discovery of the integer quantum Hall effect in 2D Chern insulators~\cite{VonKlitzingNobel}, a wide variety of topological response effects have been theoretically proposed and experimentally verified in systems with higher symmetry or dimensionality, including TIs and TCIs [Fig.~\ref{fig:response}].  Unlike in Chern insulators, the hybrid-Wannier winding in TIs and TCIs -- known as spectral flow because of its correspondence to the infinite ``flowing'' connectivity of topological boundary states~\cite{Fidkowski2011} -- is enforced by additional symmetries.  Specifically, when combinations of spatial and $\mathcal{T}$ symmetries are relaxed in TIs and TCIs, the spectral flow may be removed (trivialized), and the bulk topology reduced to that of a fragile TI or an (obstructed) atomic limit.

\subsubsection{Wilson Loops}
\label{sec:subsubWilson}

Numerous, equivalent methods have been introduced to diagnose the topology of TIs and TCIs~\cite{CharlieTI,KaneMeleZ2,AndreiTI,TRPolarization,AlexeyVDBWannier,QHZ,MooreBalents,CohVDBChern,MooreAFMTI,SSGInvariant,FulgaAnon}.  However, most methods cannot be generalized to realistic models of real-material band structures with varying symmetries, complicated dispersion, and large numbers of occupied bands.  To date, the most consistently reliable and generalizable means of diagnosing single-particle band topology is the Wilson loop method discussed earlier in this section~\cite{Fidkowski2011,TRPolarization,AlexeyVDBWannier,AndreiXiZ2,AlexeyWilson,ArisInversion,BarryFragile,Cohomological,HourglassInsulator,DiracInsulator} [Fig.~\ref{fig:Wilson}(b)].  When calculated over an energetically isolated set of bands in two or more dimensions, the Wilson loop eigenvalues indicate the ${\bf k}_{\parallel}$-dependent locations $r_{\perp}$ of the hybrid Wannier orbitals.  Hence, the appearance of a winding Wilson spectrum either indicates stable spectral flow or fragile topology.  As discussed earlier in this section, a nested Wilson loop method~\cite{WladTheory,multipole,HOTIBernevig,WiederAxion,TMDHOTI,HingeSM} was also recently introduced to calculate higher-order (hinge) spectral flow in three or more dimensions, which we further detail in Sec.~\ref{sec:higherOrder}.

\begin{figure*}[!t]
\centering
\includegraphics[width=\textwidth]{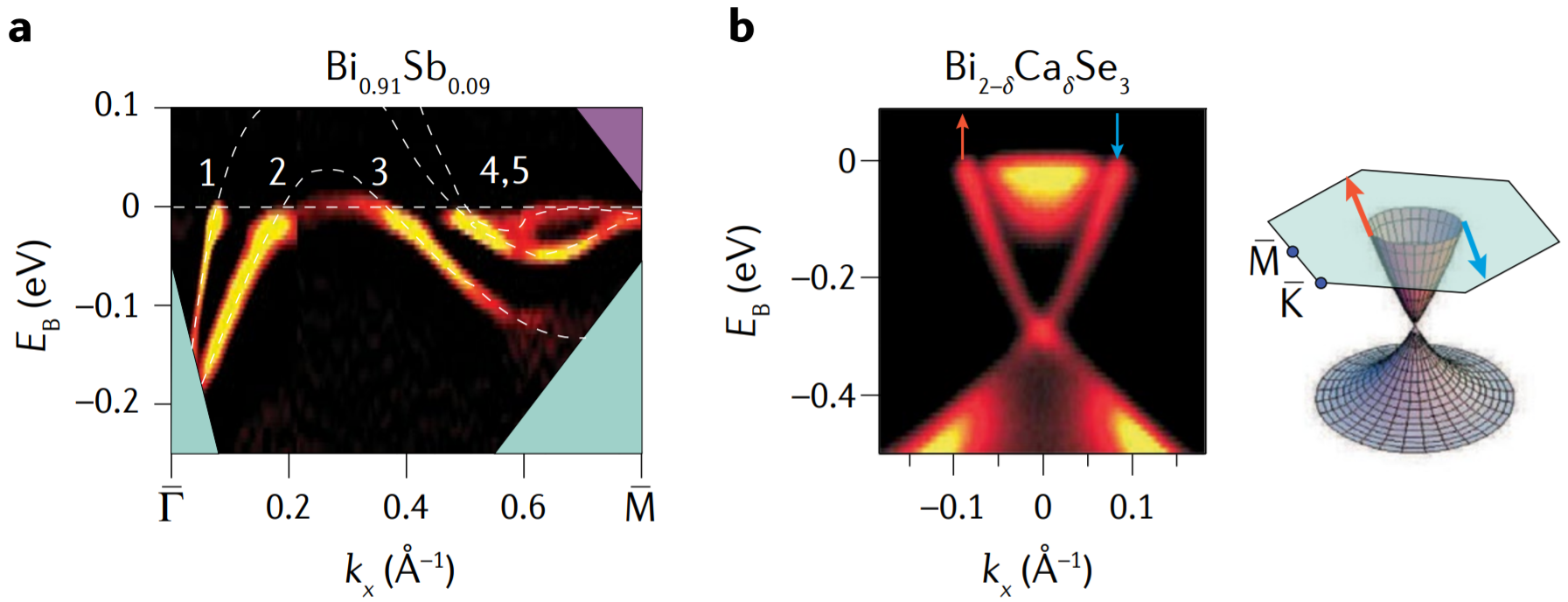}
\caption{The twofold Dirac-cone surface states of 3D topological insulators.  (a) The dispersion of the surface states in Bi$_{1-x}$Sb$_{x}$ measured through angle-resolved photoemission spectroscopy (ARPES) experiments as the electron binding energy ($E_{\rm B}$) at varying $k_{x}$~\cite{BiSbScienceKane}.  The surface bands cross the Fermi level (horizontal dashed white line) an odd number of times, indicating the spectral flow of a 3D topological insulator, as predicted in~\cite{FuKaneMele,FuKaneInversion,QHZ} (reproduced with permission from~\cite{BiSbScienceKane}).  (b, left panel) The unpaired Dirac-cone surface state of Ca-doped Bi$_2$Se$_3$ [\icsdweb{42545}, SG 166 ($R\bar{3}m$)] measured by spin-resolved ARPES.  (b, right panel) A schematic depicting the time-reversal-symmetric spin texture of the topological Dirac-cone surface states (reproduced with permission from~\cite{TunableTIZahid}).}
\label{fig:3DTI}
\end{figure*}

Because the (nested) Wilson loop eigenvalues, when calculated over bands that are not fragile topological, correspond to the surface (hinge) spectrum of a crystal~\cite{Fidkowski2011,TRPolarization,AlexeyVDBWannier,AndreiXiZ2,AlexeyWilson,ArisInversion,BarryFragile,multipole,WladTheory,HOTIBernevig,Cohomological,HourglassInsulator,DiracInsulator,WiederAxion,TMDHOTI,HingeSM}, Wilson-loop calculations have emerged as a highly accurate means of identifying and classifying the stable topology of real materials.  Implementing the formulations in~\cite{Fidkowski2011,AndreiXiZ2,AlexeyWilson}, researchers have released publicly available packages for calculating the Wilson loops of tight-binding models~\cite{PythTB} and \emph{ab-initio} electronic structures~\cite{Z2pack}.  However, the Wilson-loop method has major drawbacks that largely prevent its automation in large-scale materials discovery.  First, to compute the Wilson loop for a 3D material, one must calculate the occupied eigenstates on a 2D manifold of the BZ using a very dense mesh of ${\bf k}$ points, which is demanding in terms of both memory and processing time.  Second, because Wilson loops are path-dependent, their implementation requires the manual input of a ${\bf k}$-path specifically chosen to diagnose a particular topological phase.  For example, to identify the glide-protected hourglass TCI phase in KHgSb [\icsdweb{56201}, SG 194 ($P6_{3}/mmc$)]~\cite{Cohomological,HourglassInsulator} and the fourfold Dirac TCI phase in Sr$_2$Pb$_3$ [\icsdweb{648570}, SG 127 ($P4/mbm$)]~\cite{DiracInsulator}, researchers implemented a complicated, glide-resolved Wilson loop that bent along perpendicular planes in the BZ.  A bent Wilson loop was also recently employed to identify Weyl TSM phases in materials with fourfold rotoinversion symmetry~\cite{S4Weyl1,S4Weyl2}.

\subsubsection{Symmetry-Based Indicators}
\label{sec:subsubSIs}

Unlike (nested) Wilson loop calculations, symmetry-based indicators of band topology (SIs)~\cite{QWZ,ChenBernevigTCI,SlagerSymmetry}, such as the Fu-Kane parity criterion~\cite{FuKaneInversion}, only require the determination of the eigenvalues of the occupied bands at high-symmetry BZ points, and are therefore computationally efficient and generalizable to large families of materials with the same symmetries.  As more elaborate TI and TCI phases have been discovered, the set of Fu-Kane-like SIs of stable topology has continued to grow.  In Sec.~\ref{sec:TQC}, we discuss how the discoveries of TQC and related methods were exploited to compute the complete SIs of the 230 nonmagnetic SGs~\cite{HOTIBernevig,HOTIBismuth,AshvinIndicators,HOTIChen,ChenTCI,AshvinTCI,ZhidaSemimetals}, and facilitated recent large-scale efforts to identify nonmagnetic topological materials~\cite{AndreiMaterials,AndreiMaterials2,ChenMaterials,AshvinMaterials1,AshvinMaterials2}.

It is important to emphasize that SIs generically provide an incomplete characterization of bulk topology.  For example, the $n$-fold-rotation SIs introduced in~\cite{ChenBernevigTCI} only specify the $\mathbb{Z}$-valued Chern number up to the integer $n$.  Additionally, there are large numbers of materials, such as noncentrosymmetric crystals, whose SGs do not carry eigenvalue indicators~\cite{AshvinIndicators,ChenTCI}.   Therefore, for many crystal structures and materials, the (nested) Wilson loop remains the only complete means of identifying the bulk topology.  Recently, for example, the higher-order TCI phases of noncentrosymmetric $\gamma$-MoTe$_2$ and $\gamma$-WTe$_2$ crystals [\icsdweb{14348}, SG 31 ($Pmn2_{1}$)] were identified through the computation of a nested Wilson loop~\cite{TMDHOTI}.

\section{The First Topological Insulators}
\label{sec:firstTIs}

Over 20 years passed between the characterization of Chern insulators and polyacetylene and the discovery of the first $\mathcal{T}$-symmetric TIs.  In a key intermediate step, Duncan Haldane in 1988 formulated a model of a Chern insulator in a crystal with a periodic -- but net-zero -- magnetic field~\cite{HaldaneModel}.  Like all Chern insulators, the Haldane model exhibits 1D chiral edge modes.  In 2004, when constructing tight-binding models of graphene~\cite{GrapheneReview}, Kane and Eugene Mele formulated a model of a 2D TI that, when the nearly vanishing effects of SOC are incorporated, reduces to two, time-reversed copies of the Haldane model.  Hence, the Kane-Mele model exhibits helical edge modes~\cite{CharlieTI}.  Crucially, Kane and Mele deduced that the helical edge modes are prevented from gapping by $\mathcal{T}$ symmetry~\cite{CharlieTI,KaneMeleZ2}.  A more realistic formulation of 2D TIs in HgTe quantum wells was introduced shortly afterwards by Andrei Bernevig, Taylor Hughes, and Shou-Cheng Zhang~\cite{AndreiTI}, and was subsequently confirmed through transport measurements~\cite{HgTeExp}.  Bernevig, Hughes, and Zhang also crucially identified band inversion as a general mechanism for obtaining topological phases~\cite{AndreiTI}.

In both the Kane-Mele and Bernevig-Hughes-Zhang models of 2D TI phases, the bulk topology can be deduced by a $\mathbb{Z}_{2}$ invariant that tracks the helical winding of the partial polarization, or the Berry phase per half of the occupied states~\cite{KaneMeleZ2,TRPolarization}.  However, in practice, the Kane-Mele $\mathbb{Z}_{2}$ invariant is exceedingly difficult to compute in realistic models, as it requires a careful treatment of matrix square roots and Pfaffians, as well as the computationally intensive process of determining, storing, and integrating the occupied states at each ${\bf k}$ point.  Though a handful of calculations based on the Kane-Mele Pfaffian invariant have been employed to diagnose the topology of TI and TCI models~\cite{MurakamiBismuth2,KaiSunPfaffian}, the numerical prediction of TI and TCI phases in real materials without SIs has primarily been accomplished through the equivalent Wilson-loop method introduced in~\cite{Fidkowski2011,AndreiXiZ2,AlexeyWilson}.  Since the prediction by Bernevig, Hughes, and Zhang of 2D TI phases in HgTe quantum wells, 2D TI phases have been predicted in monolayers of jacutingaite (Pt$_2$HgSe$_3$)~\cite{Marzari2DQSHjacutingaite}, predicted and experimentally explored in monolayers of bismuth~\cite{MurakamiBismuth1,MurakamiBismuth2,AliEarlyBismuth,BismutheneSiCMonoExp} and monolayers of the transition-metal dichalcogenide (TMD) MoS$_2$~\cite{MoS2MonoTI}, and have been predicted and experimentally confirmed in monolayers of WTe$_2$~\cite{LiangFuTMD,SanfengWTe2,WTe2STMARPESNatPhys} and Ta$_2$Pd$_3$Te$_5$~\cite{ZhijunNewQSHTheory,ZhijunNewQSHExp}.  Notably, Pt$_2$HgSe$_3$ realizes a weak TI phase when stacked into 3D structures~\cite{Marzari3DWTIjacutingaiteTheory,Fulga3DWTIjacutingaiteTheory,Marzari3DWTIjacutingaiteExp}, as we will shortly discuss below, whereas bismuth~\cite{HOTIBismuth} and the TMDs MoTe$_2$ and WTe$_2$~\cite{TMDHOTI,AshvinMaterials1,AshvinMaterials2} become higher-order TCIs with 1D helical hinge modes when stacked into 3D.

The foundation for topological material discovery in 3D solid-state materials was most firmly established in 2006, when several independent groups predicted the existence of 3D TIs.  In 3D TIs, the bulk topology is characterized by a 3D $\mathbb{Z}_{2}$ invariant, and each surface exhibits an unpaired (anomalous) twofold Dirac cone~\cite{FuKaneMele,FuKaneInversion,MooreBalents,QHZ,Roy3DTI}.  In particular, Fu and Kane crucially recognized that if a 3D TI is also symmetric under the exchange of spatial coordinates -- known as spatial inversion ($\mathcal{I}$) or parity symmetry -- then the $\mathbb{Z}_{2}$ invariant can simply be deduced from the bulk parity eigenvalues of the occupied bands at the $\mathcal{T}$-invariant ${\bf k}$ (TRIM) points~\cite{FuKaneInversion}.  Because the integral formulation of the 3D $\mathbb{Z}_{2}$ invariant is even more mathematically and computationally intensive than its 2D counterpart~\cite{FuKaneMele,FuKaneInversion,QHZ}, it provided little help towards predicting 3D TI phases in real materials.  Conversely, the Fu-Kane parity criterion immediately provided a simple recipe for discovering centrosymmetric 3D TIs: generalizing the earlier link between TI phases and band inversion recognized by Bernevig, Hughes, and Zhang, Fu and Kane predicted 3D TI phases in insulating compounds with odd numbers of band inversions at TRIM points between bands with opposite parity eigenvalues.  Fu, Kane, and Mele then specifically proposed a band-inversion-driven 3D TI in Bi$_{1-x}$Sb$_{x}$ alloys, whose characteristic anomalous surface states were shortly afterwards measured in experiment~\cite{CavaHasanTI,BiSbScienceKane} [Fig.~\ref{fig:3DTI}(a)].  In the following years, more idealized 3D TI phases were identified in Bi$_2$Se$_3$ [\icsdweb{42545}, SG 166 ($R\bar{3}m$)], Bi$_2$Te$_3$ [\icsdweb{617192}, SG 166 ($R\bar{3}m$)], Sb$_2$Te$_3$ [\icsdweb{193346}, SG 166 ($R\bar{3}m$)], and Bi$_{2-\delta}$Ca$_{\delta}$Se$_3$ crystals~\cite{BiSeXia,BiTeHsieh,BiTeSCZ,TunableTIZahid,Sb2Te3JunctionExp,Sb2Te3SurfaceSC} [Fig.~\ref{fig:3DTI}(b)].  These 3D TI phases were initially observed in angle-resolved photoemission spectroscopy (ARPES) experiments~\cite{ARPESReviewYulin,ARPESReviewHongDing}, and anomalous Dirac-cone surface states in 3D TIs were subsequently confirmed through quasiparticle interference in scanning tunneling microscopy (STM) experiments~\cite{YazdaniQPI3DTI}.

\begin{figure*}[!t]
\centering
\includegraphics[width=\textwidth]{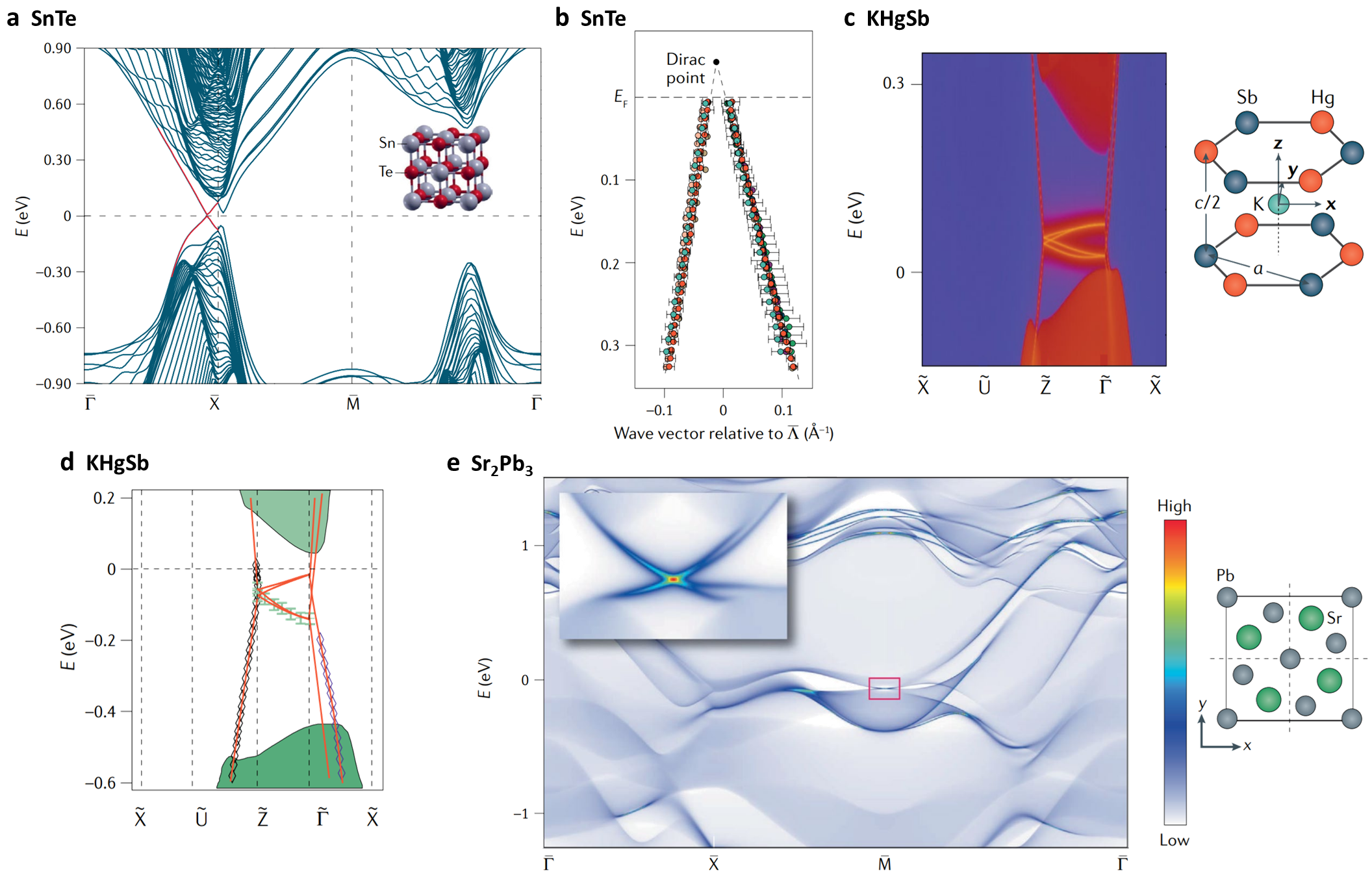}
\caption{3D topological crystalline insulators.  (a) A tight-binding calculation predicting the presence of twofold Dirac-cone surface states (crossing red bands) in the mirror topological crystalline insulator (TCI) SnTe [\icsdweb{601065}, SG 225 ($Fm\bar{3}m$)]~\cite{HsiehTCI,SnTeExp} (reproduced with permission from~\cite{HsiehTCI}).  (b) Angle-resolved photoemission spectroscopy (ARPES) data for various photon energies confirming the existence of TCI surface Dirac cones in SnTe, plotted along a line of surface mirror symmetry ($\bar{\Gamma}\bar{X}$) and centered at the surface Dirac cone at $k_{x}=\bar{\Lambda}$ (reproduced with permission from~\cite{SnTeExp}).  The error bars in (b) reflect the uncertainties originating from the momentum resolution and the standard deviation in the peak positions in the ARPES data.  (c) Topological hourglass surface fermions in the glide TCI KHgSb [\icsdweb{56201}, SG 194 ($P6_{3}/mmc$)]~\cite{Cohomological,HourglassInsulator,HourglassObserve,ZeroHallExp}, calculated from first principles (reproduced with permission from~\cite{HourglassInsulator}).  (d) ARPES data supporting the existence of topological hourglass surface states in KHgSb (reproduced with permission from~\cite{HourglassObserve}).  (e)  Anomalous fourfold Dirac-cone surface states in the nonsymmorphic Dirac TCI Sr$_2$Pb$_3$ [\icsdweb{648570}, SG 127 ($P4/mbm$)], calculated from first principles (reproduced with permission from~\cite{DiracInsulator}).  The color bar in (e) indicates the computed surface spectral weight in arbitrary units, which is high for the fourfold Dirac cone in the red rectangle at $\bar{M}$ near $E=0$ (shown in closer detail in the inset panel).}
\label{fig:TCI}
\end{figure*}

Lastly, concurrently with the prediction of 3D TIs, researchers also proposed the existence of weak 3D TIs (WTIs).  From a theoretical perspective, WTIs can be deformed into position-space stacks of 2D TIs that contain one 2D TI layer in each stack unit cell, where the normal vector of each 2D TI layer is oriented along the $(Z_{2,1}\hat{\bf x} + Z_{2,2}\hat{\bf y} + Z_{2,3}\hat{\bf z})$-direction (the stacking direction)~\cite{FuKaneMele,FuKaneInversion,MooreBalents,Roy3DTI}.  The three $\mathbb{Z}_{2}$-valued indices $Z_{2,i}$ are termed the weak indices, and the stacking direction $(Z_{2,1},Z_{2,2},Z_{2,3})$ is termed the weak-index vector.  WTIs owe their name to the recognition that in samples with even numbers of layers, all of the 2D TI helical modes may pairwise annihilate, rendering the WTI sample devoid of topological boundary states~\cite{AdyWeak}.  Hence, few-layer WTI samples are expected to exhibit a layer-dependent 2D quantum spin Hall effect~\cite{FuKaneMele,FuKaneInversion,MooreBalents,Roy3DTI,AdyWeak,HalfQSHSCZWTI}.  WTI phases have been predicted and experimentally confirmed through ARPES probes in BiI [\icsdweb{1559}, SG 12 ($C2/m$)]~\cite{FanBiBrWTI,BiIWTIExp}, Pt$_2$HgSe$_3$ [\icsdweb{185808}, SG 164 ($P\bar{3}m1$)]~\cite{Marzari3DWTIjacutingaiteTheory,Fulga3DWTIjacutingaiteTheory,Marzari3DWTIjacutingaiteExp}, and strained ZrTe$_5$ [\icsdweb{85507}, SG 63 ($Cmcm$)]~\cite{ZrTe5Prediction,ZrTe5WTIExp}.  A WTI phase was also engineered in a graphene-like layered structure of Bi$_{14}$Rh$_3$I$_9$~\cite{EngineeredWTIGrowth}, as confirmed through bulk ARPES probes~\cite{EngineeredWTIARPES} and STM experiments demonstrating the presence of step-edge helical modes~\cite{EngineeredWTISTM}.  Most recently, a WTI phase was identified through first-principles and ARPES investigations~\cite{SlagerWTI} of a relatively unstudied structure of RhBi$_2$ that is not recorded in the ICSD~\cite{ICSD1,ICSD2,RuckSlagerWTI}.

\section{Topological Crystalline Insulators}
\label{sec:TCIs}

When searching for candidate 3D TIs, Fu and Kane initially suggested that Pb$_{1-x}$Sn$_x$Te alloys exhibited a 3D TI phase separated by two equally trivial insulators: PbTe [\icsdweb{648585}, SG 225 ($Fm\bar{3}m$)] and SnTe [\icsdweb{601065}, SG 225 ($Fm\bar{3}m$)]~\cite{FuKaneInversion}.  However, it was quickly realized~\cite{HsiehTCI}, and subsequently confirmed through ARPES experiments~\cite{SnTeExp} [Fig.~\ref{fig:TCI}(a,b)], that SnTe is in fact not a trivial insulator.  Rather, SnTe was recognized to be a new variant of 3D TI -- a 3D TCI -- in which the Fu-Kane-Mele 3D $\mathbb{Z}_{2}$ invariant is trivial, and the bulk stable topology is instead protected by mirror symmetry (as well as bulk fourfold rotation symmetry~\cite{HOTIBernevig,HOTIChen,ChenRotation}).  In SnTe, there are even numbers of twofold Dirac-cone surface states lying along surface lines of mirror symmetry~\cite{HsiehTCI}.  More broadly, in TCIs, the bulk topology is enforced by specific combinations of unitary symmetries, such as mirror reflection, and antiunitary symmetries, such as $\mathcal{T}$.  Anomalous topological surface states consequently manifest on crystal facets whose 2D symmetry (wallpaper) groups include the projections of the 3D symmetries that protect the bulk topology~\cite{LiangTCI,TeoFuKaneTCI,DiracInsulator,ChaoxingWallpaper,SlagerNatPhysDisorder}.

After the discovery of mirror TCIs, researchers proposed nonsymmorphic TCI phases protected by bulk glide reflection~\cite{SSGMobius,ChenMobius,Cohomological,HourglassInsulator,DiracInsulator}.  In magnetic M{\"o}bius TCIs~\cite{SSGMobius,ChenMobius}, unpaired Dirac-cone surface states -- like those of 3D TIs -- appear along surface glide lines.  In $\mathcal{T}$-symmetric hourglass TCIs~\cite{Cohomological,HourglassInsulator,DiracInsulator} -- which are topologically equivalent to time-reversed, superposed pairs of M{\"o}bius TCIs -- the surface bands cross in hourglass-like connectivities along surface glide lines, forming pairs of twofold Dirac cones [Fig.~\ref{fig:TCI}(c,d)].  Most recently, an exotic, nonsymmorphic Dirac insulator phase with an anomalous fourfold surface Dirac cone -- twice the degeneracy of a 3D TI surface state -- was proposed in crystals with two perpendicular, surface-projecting glide planes [Fig.~\ref{fig:TCI}(e)]~\cite{DiracInsulator}.  An hourglass TCI phase was predicted~\cite{Cohomological,HourglassInsulator} and incipiently confirmed through ARPES~\cite{HourglassObserve} and quantum oscillations~\cite{ZeroHallExp} in KHgSb [\icsdweb{56201}, SG 194 ($P6_{3}/mmc$)] [Fig.~\ref{fig:TCI}(c,d)].  Shortly afterwards, an hourglass TCI phase was also predicted in Ba$_5$In$_2$Sb$_6$ [\icsdweb{62305}, SG 55 ($Pbam$)], and a nonsymmorphic Dirac TCI phase was predicted in Sr$_2$Pb$_3$ [\icsdweb{648570}, SG 127 ($P4/mbm$)] [Fig.~\ref{fig:TCI}(e)]~\cite{DiracInsulator}.  From a bulk perspective, it is important to note that whereas centrosymmetric mirror TCIs may be identified through SIs~\cite{ChenBernevigTCI,AshvinIndicators,ChenTCI,AshvinTCI}, nonsymmorphic TCIs may only be identified by performing a Wilson loop or equivalent calculation~\cite{SSGMobius,ChenMobius,Cohomological,HourglassInsulator,DiracInsulator,ChenTCI}.

Recently, researchers have also discovered the existence of 3D higher-order TCIs (HOTIs), which are characterized by gapless 1D hinge states that sometimes coexist with 2D surface Dirac cones.  Specifically, although the lowest-symmetry HOTI models exhibit gapped 2D surfaces and gapless 1D hinges~\cite{AshvinIndicators,ChenTCI,AshvinTCI,TMDHOTI}, typical solid-state HOTIs have additional crystal symmetries that can protect gapless 2D surface states.  Hence, classifying a typical solid-state TCI as a HOTI or as a first-order TCI is in practice a matter of preference: depending on the sample termination, the same crystal may exhibit topological surface states or gapped facets separated by topological hinge states.  For example, the first observed mirror TCI -- SnTe -- has recently been reclassified as a HOTI~\cite{HOTIBernevig,ChenTCI,AndreiMaterials,ChenMaterials,AshvinMaterials1,AshvinMaterials2}.  Specifically, when bulk or surface mirror symmetry is relaxed in SnTe by strain or sample termination while preserving fourfold rotation symmetry, the surface states are no longer protected by mirror symmetry (though some surface states remain protected by the product of twofold rotation and $\mathcal{T}$ symmetry~\cite{ChenRotation}).  In SnTe samples with broken mirror symmetry, the hinges between facets with gapped surface states are predicted to bind 1D helical modes~\cite{HOTIBernevig,ChenTCI,HOTIChen,ChenRotation}.  Furthermore, first-principles calculations have suggested that PbTe [\icsdweb{648588}, SG 225 ($Fm\bar{3}m$)] -- which is isostructural to SnTe -- may also realize TCI and HOTI phases depending on sample quality and preparation details~\cite{AndreiMaterials,ChenMaterials,AshvinMaterials1,AshvinMaterials2,BarryPbTe}.  Lastly, other TCIs phases -- most notably including the nonsymmorphic Dirac TCIs -- also represent the parent phases of HOTIs with helical hinge states~\cite{DiracInsulator,TMDHOTI,JenHOTI,MaiaCICEDiracInsulatorHOTI}.

\section{Topological Semimetals}
\label{sec:TSMs}

Historically, the term semimetal has been used to describe solid-state materials in which there is a small but nonzero density of states at $E_{F}$~\cite{AshcroftMermin}.  However, there are two distinct mechanisms for realizing a semimetallic density of states.  In the first case, such as in the archetypal semimetal bismuth [\icsdweb{64703}, SG 166 ($R\bar{3}m$)]~\cite{LiuAllenBismuth}, there is a gap at all ${\bf k}$ points, but the system remains metallic due its negative indirect band gap.  As emphasized by Fu and Kane~\cite{FuKaneInversion}, bismuth-like semimetals may still exhibit nontrivial band topology owing to the presence of an energy gap between the valence and conduction manifolds at all ${\bf k}$ points.  Hence in more modern literature, semimetals like bismuth are referenced as band insulators.  In recent years, bismuth itself was in fact recognized to be a higher-order topological crystalline band insulator~\cite{HOTIBismuth}.

In the second case, a semimetallic density of states also occurs in crystalline solids in which the bulk bands meet at $E_{F}$ in only a handful of nodal points, but are otherwise separated by an energy gap.  Such systems are termed nodal semimetals.  The prototypical 2D nodal semimetal phase occurs in graphene, in which the spin-degenerate bands meet in only two, linearly dispersing points~\cite{GrapheneReview,GrapheneDirac,semenoffDirac,meleDirac}.  If the bands in a $d$-dimensional [$d$-D] nodal semimetal exhibit nontrivial $(d-1)$-D or $(d-2)$-D polarization, fragile, or stable topology, then the nodal semimetal is further classified as a topological semimetal [TSM].  For example in zigzag-terminated graphene, the edge ${\bf k}$ points between the projections of the bulk nodal points exhibit nearly-flat bands enforced by bulk polarization topology~\cite{GrapheneEdgeMullen,GrapheneEdge1,GrapheneEdge2,GrapheneEdgeFan}; hence, graphene is the simplest TSM.  Because the low-energy mathematical description of the nodal points in graphene (the ${\bf k}\cdot {\bf p}$ Hamiltonian) resembles the Dirac equation for relativistic particles~\cite{semenoffDirac,meleDirac}, the nodal points were termed 2D Dirac points.  As we discuss in Sec.~\ref{sec:DiracWeyl}, in 3D TSMs, there can exist bulk nodal degeneracies whose ${\bf k}\cdot{\bf p}$ Hamiltonians resemble the high-energy Dirac and Weyl equations; hence, in these 3D TSMs, the nodal points are respectively termed 3D Dirac and Weyl points.  Initially, the search for TSMs and TIs was motivated as an effort to observe condensed-matter realizations of high-energy phenomena.  For example, the surface states of 3D TIs are unpaired twofold Dirac cones, representing half of the degeneracy of each of the Dirac points in graphene~\cite{FuKaneMele,FuKaneInversion,QHZ,CavaHasanTI,FanTISurface}.  From the perspective of high-energy theory, the unpaired Dirac cones on 3D TI surfaces represent experimentally observable topological exceptions to the parity anomaly: a fermion doubling theorem first formulated in the context of high-energy models that requires symmetry-stabilized twofold Dirac cones to appear in pairs in 2D systems~\cite{WittenParity,RedlichParity,JackiwParity}.  However, as we discuss in Sec.~\ref{sec:chiralCrystals}, researchers have since discovered numerous 3D solid-state TSMs in which the nodal degeneracies at $E_{F}$ are neither Dirac nor Weyl fermions.

\vspace{-0.1in}
\subsection{Dirac and Weyl Semimetals}
\label{sec:DiracWeyl}

For a time, the analogy between solid-state and high-energy physics held relatively well in 3D semimetals.  Searching for 3D generalizations of graphene, researchers identified 3D Dirac semimetals through two, parallel tracks.  First, researchers recognized that nodal fermions at high-symmetry ${\bf k}$ points transform in small coreps of the little group $G_{\bf k}$~\cite{manes,SteveDirac}.  Because the small coreps of the little groups in all 230 nonmagnetic SGs were enumerated in~\cite{BigBook}, it was possible to target specific ${\bf k}$ points in specific SGs for high-symmetry-point, essential 3D Dirac fermions: fourfold degeneracies with linear dispersion in all three directions that transform in four-dimensional, double-valued small coreps at high-symmetry BZ points.  More broadly, essential nodal points were recognized to be features of the minimal band connectivity in each SG, and can also lie along high-symmetry lines and planes in nonsymmorphic and non-primitive SGs~\cite{WiederLayers,WPVZ,WPVZprl}.  Searching for high-symmetry-point, essential Dirac fermions, researchers theoretically predicted 3D Dirac semimetal phases in hypothetical, metastable structural phases of the BiO$_2$~\cite{SteveDirac} and BiZnSiO$_4$ families~\cite{JuliaDirac}.  Unfortunately, the structural phases of BiO$_2$ and BiZnSiO$_4$ in~\cite{SteveDirac,JuliaDirac} have thus far proved to be experimentally inaccessible.  Nevertheless, high-symmetry-point, essential nodal fermions have since been discovered in a wealth of readily accessible materials, validating the methods introduced in~\cite{manes,SteveDirac}.  Furthermore,~\cite{SteveDirac,JuliaDirac} represented some of the earliest attempts at combining crystal symmetry, group theory, and material databases for topological materials discovery, laying the foundation for large-scale materials discovery efforts.

Other researchers hunted for solid-state realizations of 3D Dirac fermions along a second, parallel research track focused on band inversion.  Unlike high-symmetry-point Dirac fermions, band-inversion 3D Dirac fermions are defined as fourfold degeneracies with linear dispersion in all three directions that form at the crossing points between two, twofold-degenerate bands that transform in different small coreps of a little group along a high-symmetry BZ line or plane.  In band-inversion semimetals, unlike in essential, high-symmetry-line or plane semimetals, the nodal points at $E_{F}$ are not required to be present in the minimal band connectivity, and are instead pairwise nucleated by band inversion.  Searching for compounds in which the bulk bands were strongly inverted along axes of fourfold or sixfold rotation symmetry, researchers predicted band-inversion 3D Dirac semimetal phases in Cd$_3$As$_2$ [\icsdweb{609930}, SG 137 ($P4_{2}/nmc$)]~\cite{ZJDirac} and Na$_3$Bi [\icsdweb{26881}, SG 194 ($P6_{3}/mmc$)]~\cite{ZJDirac2}, which were concurrently confirmed in ARPES experiments~\cite{CavaDirac1,CavaDirac2,ZJDiracExp1,NaDirac}.  Tilted (type-II) variants of band-inversion Dirac TSM phases have also been proposed in the VAl$_3$ [\icsdweb{167811}, SG 139 ($I4/mmm$)]~\cite{Type2Dirac1} and YPd$_2$Sn [\icsdweb{105699}, SG 225 ($Fm\bar{3}m$)]~\cite{Type2Dirac2} material families, and observed in PtTe$_2$ [\icsdweb{649753}, SG 164 ($P\bar{3}m1$)] through ARPES experiments~\cite{Type2DiracPtTe2ARPES}.

During their initial characterization, the designation of Dirac semimetals as TSMs was controversial.  Specifically, on the one hand, it was recognized that Dirac semimetals could exhibit arc-like surface states~\cite{SYDiracSurface,ZJSurface,AshvinDiracSurface,SchnyderDirac}, and that the same band inversions that create Dirac points can also result in high-symmetry BZ planes exhibiting the same stable bulk topology and surface states as 2D TIs and TCIs~\cite{NagaosaDirac}.  Indeed, arc-like surface states were experimentally observed in Dirac semimetal phases in Cd$_3$As$_2$, Na$_3$Bi, and PtSn$_4$ [\icsdweb{54609}, SG 68 ($Ccce$)] through ARPES~\cite{ZJSurface,SYDiracSurface,KaminskiDiracArc}, quantum oscillations~\cite{AshvinDiracSurface,AnalytisOscillation}, and quasiparticle interference in STM~\cite{AndreiAliDiracQPI,AshvinAliDiracQPI}.  However, on the other hand, it was also recognized that the arc-like states were not topological~\cite{KargarianDiracArc1}, and that not all Dirac semimetals are required to exhibit surface states~\cite{KargarianDiracArc2,ThomaleArc}.  Recently, it was discovered that large families of Dirac semimetals, including Cd$_3$As$_2$, exhibit higher-order hinge states as a direct topological consequence of their bulk Dirac points, definitely resolving the status of Dirac semimetals as TSMs~\cite{HingeSM}.

In addition to 3D Dirac fermions, researchers also discovered solid-state realizations of conventional Weyl fermions, defined as twofold degeneracies with linear dispersion in all three directions~\cite{MurakamiWeyl,AshvinWeyl,BurkovBalents}.  Unlike Dirac fermions, conventional Weyl fermions carry a nontrivial topological invariant in the absence of symmetry -- the Chern number $|C|=1$ evaluated on a sphere around the Weyl point (also known as the chiral charge of the Weyl point) -- and hence can form at any point in the BZ of a crystal with singly-degenerate bands, even if the crossing bands are labeled with the same small coreps.  More generally, Weyl points are condensed-matter realizations of chiral fermions: nodal degeneracies with nontrivial, $\mathbb{Z}$-valued monopole (chiral) charges reflecting the difference in the Chern number evaluated in 2D BZ planes above and below the chiral fermion.  On the 2D facets of finite crystals with Weyl points at $E_{F}$, the projections of BZ planes with nontrivial Chern numbers host topological surface Fermi arcs that span the projections of the bulk chiral fermions~\cite{AshvinWeyl}.  A conventional Weyl semimetal is therefore the simplest variant of a chiral TSM: a gapless phase of matter in which some or all of the bulk Fermi pockets carry nontrivial chiral charges.  Chiral TSMs can hence be contrasted with achiral TSMs -- such as Dirac semimetals -- in which none of the bulk Fermi pockets originate from chiral fermions with uncompensated chiral charges.

The first Weyl semimetals were predicted in magnetic phases of pyrochlore iridates like Y$_2$Ir$_2$O$_7$ [\icsdweb{187534}, SG 227 ($Fd\bar{3}m$)]~\cite{AshvinWeyl}, though Weyl points have not yet been observed in these materials.  Shortly after the discovery of 3D Dirac semimetals, band-inversion Weyl-semimetal phases were predicted~\cite{AndreiWeyl,HasanWeylDFT} and experimentally confirmed through ARPES~\cite{SuyangWeyl,LvWeylExp,YulinWeylExp} and quasiparticle interference in STM~\cite{AliWeylQPI} in the transition-metal monophosphides TaAs [\icsdweb{611451}, SG 109 ($I4_{1}md$)], TaP [\icsdweb{648185}, SG 109 ($I4_{1}md$)], NbAs [\icsdweb{16585}, SG 109 ($I4_{1}md$)], and NbP [\icsdweb{81493}, SG 109 ($I4_{1}md$)].  Type-II (tilted) Weyl points were also predicted to lie near $E_{F}$ in the noncentrosymmetric, isostructural TMDs $\gamma$-WTe$_2$ [\icsdweb{14348}, SG 31 ($Pmn2_{1}$)]~\cite{AlexeyType2} and $\gamma$-MoTe$_2$~\cite{ZJType2}, though later research~\cite{TMDHOTI} has suggested that 3D TMDs may also be HOTIs.  Higher-charge Weyl points stabilized by rotation symmetries have additionally been proposed in SrSi$_2$ [\icsdweb{24145}, SG 212 ($P4_{3}32$)]~\cite{ZahidMultiWeylSrSi2}, NbSi$_2$ [\icsdweb{601659}, SG 180 ($P6_{2}22$)]~\cite{StepanMultiWeyl}, and in the ferromagnetic phase of HgCr$_2$Se$_4$ [\icsdweb{626175}, SG 227 ($Fd\bar{3}m$)]~\cite{AndreiMultiWeyl,XiDaiMultiWeyl}.

\begin{figure*}[!t]
\centering
\includegraphics[width=\textwidth]{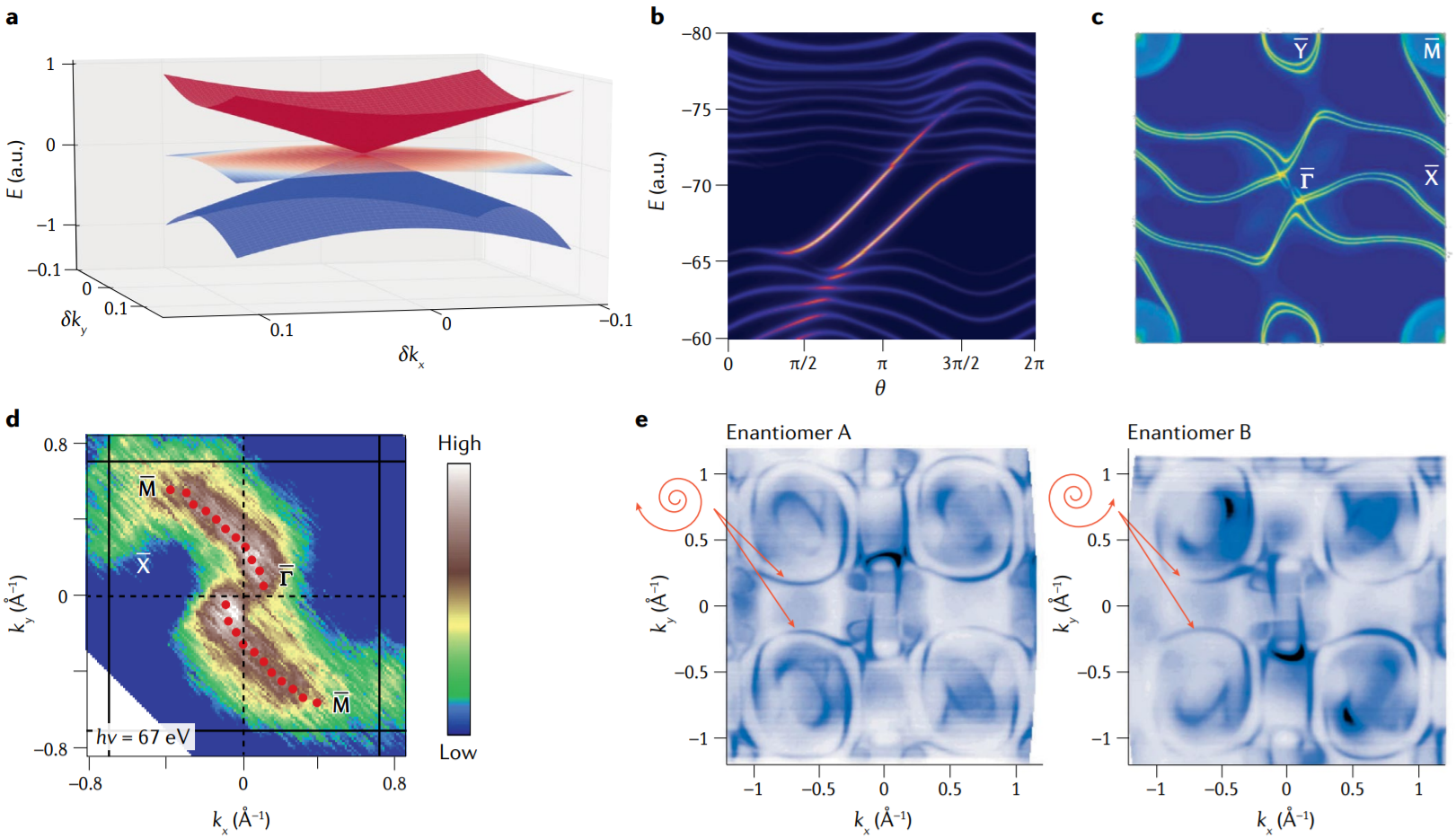}
\caption{Chiral topological semimetal phases in structurally chiral crystals.  (a) The dispersion relation of a spin-1 chiral fermion (reproduced with permission from~\cite{NewFermions}).  Because the nodal point of the spin-1 fermion in (a) is threefold-degenerate, the spin-1 fermion is an example of an unconventional quasiparticle beyond condensed-matter realizations of Dirac and Weyl fermions, which respectively correspond to fourfold and twofold nodal degeneracies with linear dispersion~\cite{DDP,NewFermions}.  (b) Surface Fermi arcs of a tight-binding model of a chiral topological semimetal with spin-1 fermions at the Fermi energy (reproduced with permission from~\cite{NewFermions}).  Two surface Fermi arcs with positive slopes cross the bulk gap in (b), indicating that the spin-1 fermions carry the chiral charges $\pm 2$.  (c) Dramatically long topological Fermi arcs spanning the 2D Brillouin zone of the surface of a 3D topological chiral crystal in the RhSi (B20) family of materials [\icsdweb{79233}, SG 198 ($P2_{1}3$)], calculated from first principles~\cite{KramersWeyl,RhSi,CoSi} (reproduced with permission from~\cite{RhSi}).  (d) Angle-resolved photoemission spectroscopy (ARPES) data confirming the presence of zone-spanning topological surface Fermi arcs in B20-CoSi [\icsdweb{260982}, SG 198 ($P2_{1}3$)]~\cite{CoSiObserveJapan,CoSiObserveHasan,CoSiObserveChina} (reproduced with permission from~\cite{CoSiObserveJapan}).  (e) ARPES data demonstrating the dependence of the helicity of the surface Fermi arcs on sample structural chirality in B20-PdGa [\icsdweb{261111}, SG 198 ($P2_{1}3$)] (reproduced with permission from~\cite{PdGaObserve}).}
\label{fig:chiralFermion}
\end{figure*}

Though most Weyl points near $E_{F}$ in solid-state materials arise from band inversion, researchers have also recently predicted essential Kramers-Weyl points pinned to TRIM and other high-symmetry points in structurally chiral crystals with relevant SOC~\cite{KramersWeyl,Andreitalk}, where structural chirality is defined through the absence of rotoinversion symmetries such as mirror or $\mathcal{I}$~\cite{FlackChirality}.  Unlike band-inversion Weyl points, Kramers-Weyl points are formed from two Bloch states at a momentum ${\bf k}$ that transform in a two-dimensional, double-valued small corep of a chiral little group $G_{\bf k}$~\cite{KramersWeyl}.  Kramers-Weyl points have since been predicted in Ag$_2$Se$_{0.3}$Te$_{0.7}$~\cite{KramersWeyl}, and have been experimentally investigated in chiral tellurium [\icsdweb{23062}, SG 152 ($P3_{1}21$)] through ARPES~\cite{ChiralTelluriumKramersWeyl} and in $\beta$-Ag$_2$Se [\icsdweb{260148}, SG 19 ($P2_{1}2_{1}2_{1}$)]~\cite{KramersWeylAg2SeExp} and few-layer chiral tellurium~\cite{FakeKramersWeylTeExp} through quantum oscillations.  The complete set of single-valued coreps corresponding to weak-SOC variants of Kramers-Weyl points was also recently enumerated~\cite{ZJPhononChiralWeyl}.  Lastly, Weyl semimetals have emerged as a playground for topological bulk response effects [Fig.~\ref{fig:response}], with the transition-metal monophosphides TaAs and NbP and the half-Heusler GdPtBi [\icsdweb{58786}, SG 216 ($F\bar{4}3m$)] in an external magnetic field notably providing platforms for the observation of the gravitational anomaly~\cite{ClaudiaGravitational,FranzReview} and the chiral anomaly~\cite{NielsenNinomiyaABJ,SonSpivakChiral,PhuonChiralDirac,ChiralAnomalyWeyl1,ChiralAnomalyWeyl2,ChiralAnomalyWeyl3,ZahidChiralAnomaly,JenChiralAnomaly}.

\subsection{Unconventional TSMs and Chiral Crystals}
\label{sec:chiralCrystals}

Following the initial characterization of Dirac and Weyl semimetals, researchers over the past five years have recognized the existence of increasingly exotic variants of unconventional TSMs without analogs in high-energy physics, including closed rings of nodal points -- known as nodal lines~\cite{YoungkukLineNode,XiLineNode,ChenWithWithout} -- and nodal points whose ${\bf k}\cdot {\bf p}$ Hamiltonians do not resemble the dispersion relations of familiar relativistic particles~\cite{DDP,NewFermions}.  In many 3D nodal-line semimetals, the surface projections of the nodal lines are filled with 2D puddles of topological, flat-band-like drumhead surface states that originate from a $\pi$ shift in the bulk polarization density (surface-directed Berry phase) between the interior and exterior of the nodal line~\cite{YoungkukLineNode,XiLineNode}.  Nodal-line TSMs with drumhead surface states have been proposed in an overwhelming number of solid-state materials~\cite{YoungkukLineNode,XiLineNode,KeeLineNode,ChenWithWithout,WiederLayers,NodalChain,NonAbelianNodalChain}, and have been experimentally investigated in numerous compounds, including Ca$_3$P$_2$ in a novel structural phase through X-ray diffraction~\cite{DiracRatPoison}; ZrSiS [\icsdweb{15569}, SG 129 ($P4/nmm$)]~\cite{LeslieLine1,LeslieLine2}, CaAgAs [\icsdweb{10017}, SG 189 ($P\bar{6}2m$)]~\cite{LineNodeExpCaAgAs}, PbTaSe$_2$ [\icsdweb{74704}, SG 187 ($P\bar{6}m2$)]~\cite{ZahidPbTaSe2}, and ZrB$_2$ [\icsdweb{615755}, SG 191 ($P6/mmm$)]~\cite{LineNodeExpZrB2} through ARPES; NbAs$_2$ [\icsdweb{81218}, SG 12 ($C2/m$)]~\cite{NodalLineOpticalConductivityNbAs2} through optical conductivity measurements; and ZrSiSe [\icsdweb{15571}, SG 129 ($P4/nmm$)] through quasiparticle interference in STM~\cite{NodalLineQPIZrSiSe}.

Next, beginning with the prediction of eightfold double Dirac points in materials including Bi$_2$AuO$_5$ [\icsdweb{82092}, SG 130 ($P4/ncc$)]~\cite{DDP} and CuBi$_2$O$_4$ [\icsdweb{56390}, SG 130 ($P4/ncc$)]~\cite{NewFermions} (though the latter would prove to be a Mott insulator due to interactions~\cite{DDPMott1,DDPMott2}), researchers identified unconventional nodal point semimetals that lie between or beyond the earlier Dirac and Weyl classification schemes [Fig.~\ref{fig:chiralFermion}(a,b)].  Examples include the band-inversion, partially dispersing threefold points predicted in $\theta$-TaN [\icsdweb{76455}, SG 187 ($P\bar{6}m2$)]~\cite{Nexus1} and in the WC family [\icsdweb{672362}, SG 187 ($P\bar{6}m2$)]~\cite{Nexus2,Nexus3} and confirmed through ARPES in MoP [\icsdweb{644091}, SG 187 ($P\bar{6}m2$)]~\cite{MoP} and WC~\cite{WCarc}, the coexisting Dirac and Weyl points predicted in the SrHgPb family [\icsdweb{602710}, SG 186 ($P6_{3}mc$)]~\cite{YoungkukDiracWeyl}, and the achiral sixfold fermions observed in PdSb$_2$ [\icsdweb{77109}, SG 205 ($Pa\bar{3}$)] through ARPES experiments~\cite{AchiralPdSb2Exp}.  In the aforementioned band-inversion threefold point and mixed Dirac-Weyl TSMs, the nontrivial bulk stable topology and topological surface states originate from a combination of band inversion and chiral-fermion (Weyl-point) descent relations, but are not directly related to the bulk threefold and Dirac points themselves~\cite{WiederTripleSummary}.  High-symmetry-point sixfold fermions were also predicted to lie near $E_{F}$ in CoSi [\icsdweb{260982}, SG 198 ($P2_{1}3$)] and AlPt [\icsdweb{672522}, SG 198 ($P2_{1}3$)]~\cite{NewFermions}.

Most notably, recent advances in TSMs have come from identifying conventional and unconventional nodal degeneracies in structurally chiral crystals.  First, the authors of~\cite{KramersWeyl} demonstrated that all point-like degeneracies in structurally chiral crystals necessarily exhibit nontrivial chiral charges, implying that all nodal-point TSMs with structurally chiral SGs are topologically chiral TSMs (semimetallic phases in which bulk Fermi pockets carry nontrivial chiral charges).  This led to the \emph{ab-initio} prediction~\cite{RhSi,CoSi} that members of the RhSi (B20) chiral crystal family [\icsdweb{79233}, SG 198 ($P2_{1}3$)] -- including the aforementioned examples of CoSi and AlPt~\cite{NewFermions} -- host topological Fermi arcs that span the surface BZ and link the projections of essential, high-symmetry-point, unconventional, chiral fermions [Fig.~\ref{fig:chiralFermion}(c)].  Zone-spanning surface Fermi arcs were subsequently observed in ARPES measurements of CoSi~\cite{CoSiObserveJapan,CoSiObserveHasan,CoSiObserveChina} [Fig.~\ref{fig:chiralFermion}(d)] and AlPt~\cite{AlPtObserve}, and through quasiparticle interference in STM probes of CoSi~\cite{CoSiQPI}.  Following those experiments, BZ-spanning topological surface Fermi arcs were observed through ARPES in numerous other topological chiral crystals, including PdGa [\icsdweb{261111}, SG 198 ($P2_{1}3$)]~\cite{PdGaObserve} [Fig.~\ref{fig:chiralFermion}(e)], PtGa [\icsdweb{635132}, SG 198 ($P2_{1}3$)]~\cite{PtGaObserve}, RhSn [\icsdweb{650381}, SG 198 ($P2_{1}3$)]~\cite{RhSnChiralityReversal}, and PdBiSe [\icsdweb{616956}, SG 198 ($P2_{1}3$)]~\cite{PdBiSeARPES}.  Lastly, chiral TSM phases in structurally chiral crystals have also been recognized as ideal systems for experimentally probing topological response effects.  Structurally chiral TSMs have been theoretically proposed as platforms for realizing topological superconductivity~\cite{KTLawMultifoldSC,TayRongKramersWeylSC}, and recent experiments have demonstrated a nearly-quantized circular photogalvanic effect~\cite{AdolphoCPGE,FlickerMultifold,RhSi,KramersWeyl,ZahidSurfacePhotoresponse} in RhSi~\cite{CPGEMooreExp} and CoSi~\cite{PennChiralCPGE2}; quantum-oscillation signatures of topological superconductivity in AuBe [\icsdweb{58396}, SG 198 ($P2_{1}3$)]~\cite{AuBeTopologicalSuperconductor}; and high-efficiency and enantioselective catalysis in AlPt, PtGa, and PdGa~\cite{ClaudiaChiralCatalysis,OtherChiralCatalysis}.

\vspace{0.2in}
\section{Topological Quantum Chemistry and Symmetry-Based Indicators}
\label{sec:TQC}

As discussed in Sec.~\ref{sec:introTopology}, if a set of bands is not Wannierizable, then the bands are either stable or fragile topological.  In 2017, researchers exploited this reasoning to formulate the theory of TQC~\cite{QuantumChemistry,JenFragile1,Bandrep1,Bandrep2,Bandrep3,BarryFragile} as a means of predicting the ${\bf k}$-space topology of a material from its ${\bf r}$-space chemistry.  In TQC, the bands of (obstructed) atomic limits transform in band coreps that are induced from Wannier orbitals that in turn transform in the coreps of the site-symmetry groups $\{G_{\bf q}\}$ of a crystal.  Specifically, each $G_{\bf q}$ is defined as the group of symmetries that return the point ${\bf q}$ to itself in the same unit cell.  TQC is based on the statement that in any SG, the bands of any (obstructed) atomic limit transform in a linear combination of elementary band coreps -- defined as the band coreps that cannot be expressed as direct sums of other band coreps~\cite{ZakBandrep1,ZakBandrep2,Bandrep1,Bandrep3}.  The coreps of all 230 nonmagnetic SGs were made available for TQC through the~\href{https://www.cryst.ehu.es/cgi-bin/cryst/programs/bandrep.pl}{BANDREP} tool on the~\webBCSshort~\cite{QuantumChemistry,Bandrep1}.  Importantly, each elementary band corep is not just a collection of small coreps at each high-symmetry ${\bf k}$ point, but also includes the (nested) Berry phases that encode the relative positions of the underlying atomic orbitals~\cite{JenFragile1,BarryFragile,HingeSM,WiederAxion}.  Because the bands that transform in elementary band coreps span the set of Wannierizable bands in each SG, then any energetically isolated set of bands that does not transform in a band corep must be stable or fragile topological.

\begin{figure*}[!t]
\centering
\includegraphics[width=\textwidth]{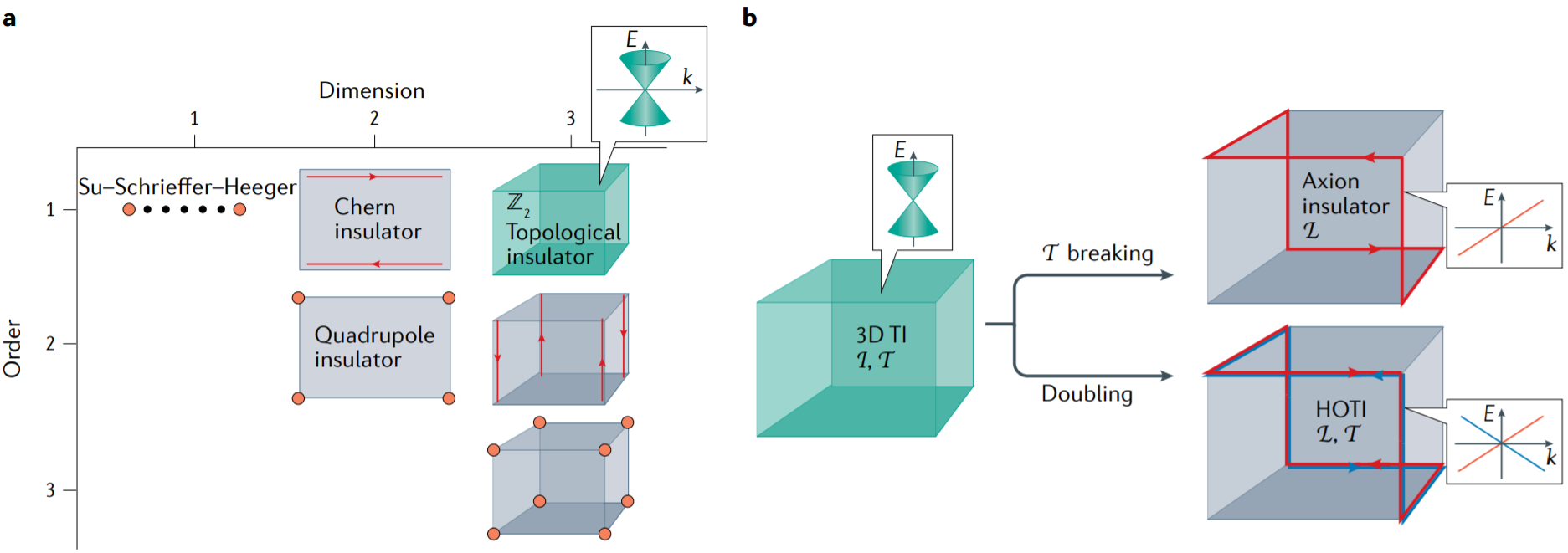}
\caption{Higher-order topological crystalline insulators.  (a) The classification of $n^\text{th}$-order topological insulators (TIs) and topological crystalline insulators (TCIs) in $d$ dimensions~\cite{WladTheory,HOTIBernevig}.  In this classification scheme, $d^\text{th}$-order TCIs in $d$ dimensions exhibit anomalous boundary charge and spin that manifest as midgap 0D states near the chiral-symmetric limit.  Hence, $d^\text{th}$-order TCIs in $d$ dimensions -- like the 2D quantized quadrupole insulator~\cite{multipole} -- are necessarily fragile topological or obstructed atomic limits~\cite{HingeSM,WiederAxion,TMDHOTI}.  We emphasize that in most 3D solid-state TCIs with 1D hinge states, like SnTe~\cite{HOTIBernevig} and bismuth crystals~\cite{HOTIBismuth,SYNewFacetBismuth}, there also exist anomalous 2D surface states~\cite{ChenRotation,WiederAxion,DiracInsulator,AshvinTCI,ChenTCI}.  Nevertheless, whether a material is interpreted to be a TCI or a higher-order TCI (HOTI), the classification of the bulk topology can unambiguously be determined through symmetry-based indicators~\cite{SlagerSymmetry,AshvinIndicators,ChenTCI,AshvinTCI,MTQC,MTQCmaterials,HOTIChen,ZhidaSemimetals,AdrianSIReview,BarryBandrepReview}, (nested) Wilson loops~\cite{Fidkowski2011,TRPolarization,AlexeyVDBWannier,AndreiXiZ2,AlexeyWilson,ArisInversion,BarryFragile,multipole,WladTheory,HOTIBernevig,Cohomological,HourglassInsulator,DiracInsulator,WiederAxion,TMDHOTI,HingeSM}, and layer constructions~\cite{HermeleSymmetry,HermeleLayerPRX,ChenTCI}.  (b) The simplest 3D HOTIs can be constructed from 3D TIs with inversion ($\mathcal{I}$) and time-reversal ($\mathcal{T}$) symmetry.  First, applying $\mathcal{I}$-symmetric magnetism to an $\mathcal{I}$- and $\mathcal{T}$-symmetric 3D TI gaps the twofold Dirac-cone surface states to reveal the intrinsic chiral hinge modes of a magnetic axion insulator~\cite{QHZ,AndreiInversion,AshvinAxion3,AshvinAxion1,AshvinAxion2,FanHOTI,FuKaneMele,FuKaneInversion,CohVDBAXI,EslamInversion,TMDHOTI,WiederAxion,VDBAxion,AshvinMagnetic,KoreanAxion,NicoDavidAXI2}.  Alternatively, superposing two $\mathcal{I}$- and $\mathcal{T}$-symmetric 3D TIs results in a non-axionic 3D HOTI phase in which unstable fourfold surface Dirac cones~\cite{DiracInsulator,TMDHOTI} gap into a network of intrinsic helical hinge modes~\cite{HOTIBismuth,ChenTCI,AshvinIndicators,AshvinTCI}.}
\label{fig:HOTI}
\end{figure*}

\vspace{0.1in}

The presence of some stable topological phases, such as Chern insulators and 3D TIs, can be inferred from the symmetry eigenvalues of the occupied bands~\cite{FuKaneInversion,QWZ,ChenBernevigTCI}.  By exploiting an enumeration of (obstructed) atomic limits equivalent to TQC, a method was developed to identify the linearly independent SI groups (such as $\mathbb{Z}_{4}\times\mathbb{Z}_{2}^{3}$) and SI formulas (such as the Fu-Kane parity criterion~\cite{FuKaneInversion}) for stable topology in each SG~\cite{SlagerSymmetry,AshvinIndicators,HOTIChen,ZhidaSemimetals}.  In this method, the SIs of an SG are determined by solving a set of Diophantine equations using the Smith-normal form~\cite{SmithForm} of the matrix of small corep multiplicities of each elementary band corep~\cite{MTQC,MTQCmaterials,AdrianSIReview,BarryBandrepReview}.  The procedure returns a set of SI topological bands that cannot be expressed as integer-valued linear combinations of elementary band coreps, and are hence stable topological.  From the SI topological bands, one may then obtain the SI groups and formulas of each SG in an arbitrary basis in which nontrivial SIs generically indicate superpositions of recognizable TIs and TCIs.  The SI groups for the 230 SGs were computed in~\cite{AshvinIndicators}.  Shortly afterwards, equivalent, intuitive bases and the associated anomalous boundary states were determined for each stable SI formula~\cite{HOTIChen,AshvinTCI,ChenTCI,ZhidaSemimetals}.  As we discuss in Sec.~\ref{sec:higherOrder}, the SI formulas in~\cite{AshvinIndicators,ChenTCI,AshvinTCI} revealed the existence of numerous symmetry-indicated HOTI phases, which included the HOTIs first identified in~\cite{HOTIBernevig,HOTIBismuth}.  SIs for fragile topological bands~\cite{AshvinFragile,JenFragile1,BarryFragile,AdrianFragile} have also been introduced~\cite{ZhidaFragile,KoreanFragileInversion}.  Most recently, the SIs of the 230 nonmagnetic SGs were used to generate large databases of topological materials~\cite{AndreiMaterials,AndreiMaterials2,ChenMaterials,AshvinMaterials1,AshvinMaterials2}.

\section{Higher-Order TCIs and TSMs}
\label{sec:higherOrder}

In 2015, a revolution in topological materials was initiated by the introduction of a model of a 2D quadrupole insulator with gapped 1D edges and anomalous, fractionally-charged 0D corner states~\cite{multipole} [Fig.~\ref{fig:HOTI}(a)].  Though earlier works had demonstrated that $d$-D topological phases could host $(d-2)$-D topological modes~\cite{TeoKaneDefect,AshvinScrewTI,FanHOTI,FranzZ4QIVortex,TeoWladProtoHOTI,DDP,SlagerDefectReview}, the quadrupole insulator model was the first system in which the presence of $(d-2)$-D boundary states was recognized to be an intrinsic consequence of the $d$-D bulk band topology.  Shortly afterwards, several works~\cite{WladTheory,HOTIBernevig, HOTIBismuth,HOTIChen,ChenRotation,EslamInversion,HigherOrderTIPiet,HigherOrderTIPiet2,DiracInsulator} demonstrated the existence of 3D HOTI phases with intrinsic 1D chiral or helical hinge states [Fig.~\ref{fig:HOTI}(b)].

The complete SIs of the 230 nonmagnetic SGs were computed in~\cite{AshvinIndicators,ChenTCI,AshvinTCI}, initially on a parallel research track to the HOTI models introduced in~\cite{multipole,WladTheory}.  However, researchers uncovered instances of models and real materials with trivial Fu-Kane parity~\cite{FuKaneInversion} and rotation-indicated TCI indices~\cite{ChenBernevigTCI}, but whose occupied bands exhibited nontrivial stable SIs~\cite{AshvinIndicators,ChenTCI,AshvinTCI}.  It was quickly recognized that the new TCI phases generically exhibited combinations of 2D surface states and 1D hinge modes, and included the HOTI phases proposed in~\cite{HOTIBernevig,HOTIChen}.  The simplest symmetry-indicated HOTI -- a ``doubled'' 3D TI -- occurs in SG 2 ($P\bar{1}$), and is formed from two superposed copies of an $\mathcal{I}$- and $\mathcal{T}$-symmetric 3D TI [Fig.~\ref{fig:HOTI}(b)].  From a physical perspective, doubled 3D TI phases occur in centrosymmetric materials with a double band inversion at the same TRIM point~\cite{HOTIBernevig,TMDHOTI}.  This realization motivated the promotion of the $\mathbb{Z}_{2}$ Fu-Kane parity criterion to a $\mathbb{Z}_{4}$ index for $\mathcal{I}$- and $\mathcal{T}$-symmetric insulators~\cite{AshvinIndicators,HOTIBernevig,TMDHOTI,ChenTCI,AshvinTCI,TMDHOTI}, coinciding with the $\mathbb{Z}_{4}$-valued SI $Z_{4}$ that results from performing for SG 2 ($P\bar{1}$) the Smith-normal form calculation outlined in Sec.~\ref{sec:TQC}.  In a concurrent work, however, it was shown that the fourfold Dirac surface states of the new phase -- the $Z_{4}=2$ doubled 3D TI -- are unstable, because the fourfold surface Dirac cones are not protected by the wallpaper group symmetries of a crystal in SG 2 ($P\bar{1}$)~\cite{DiracInsulator}.  In rapid succession, several groups determined that when a doubled 3D TI is cut in an $\mathcal{I}$-symmetric geometry, the mass of the fourfold surface Dirac cone has opposite signs on surfaces with $\mathcal{I}$-related Miller indices, resulting in a network of sample encircling, intrinsic, helical hinge modes~\cite{HOTIBismuth,ChenTCI,AshvinTCI} [Fig.~\ref{fig:HOTI}(b)].

Using the topological invariants introduced in~\cite{HOTIBernevig}, a HOTI phase protected by $\mathcal{T}$ and fourfold rotation symmetry~\cite{ChenRotation,HOTIChen} was predicted in strained SnTe [\icsdweb{601065}, SG 225 ($Fm\bar{3}m$)].  Shortly afterwards, using SIs, TQC, and reinterpreting earlier data from STM~\cite{AliEarlyBismuth} and superconducting transport experiments~\cite{BismuthSawtooth}, the existence of a $Z_{4}=2$ HOTI phase with helical 1D channels was theoretically predicted and experimentally confirmed in rhombohedral bismuth crystals [\icsdweb{64703}, SG 166 ($R\bar{3}m$)]~\cite{HOTIBismuth}.  Though subsequent works~\cite{BismuthScrewProposal,JenDefect} have provided alternative interpretations for the STM data, bismuth-like supercurrent oscillations have since been observed in other candidate HOTIs~\cite{TMDHOTI}, including MoTe$_2$ [\icsdweb{14349}, SG 11 ($P2_{1}/m$)]~\cite{PhuanOngMoTe2Hinge} and WTe$_2$ [\icsdweb{14348}, SG 31 ($Pmn2_{1}$)]~\cite{MazWTe2Exp,WTe2HingeStep,OtherWTe2Hinge}.  Additionally, hinge-state-like 1D channels have been observed in STM probes of MoTe$_2$ samples with coexisting structural orders~\cite{DavidMoTe2Exp} and in spectroscopic probes~\cite{BiBrHOTIExp,BiBrHingeExp1,BiBrHingeExp2} of the candidate HOTI BiBr [\icsdweb{1560}, SG 12 ($C2/m$)]~\cite{SYBiBr,BiBrFanHOTI,AshvinMaterials2}.  Notably, whereas the HOTI phases in bismuth, $\beta$-MoTe$_2$, and BiBr are indicated by $Z_{4}=2$~\cite{SYBiBr,BiBrFanHOTI,AshvinMaterials2}, the HOTI phases in noncentrosymmetric $\gamma$-MoTe$_2$ and $\gamma$-WTe$_2$ can only be diagnosed by computing a nested Wilson loop~\cite{TMDHOTI}.

The discovery of HOTIs also resolved previous ambiguities in the characterization of magnetically doped 3D TIs.  It had been known since the identification of the first 3D TIs that breaking $\mathcal{T}$ symmetry can drive a 3D TI into an axion insulator (AXI) phase with gapped bulk and surface states and gapless (chiral) hinge modes~\cite{QHZ,AndreiInversion,AshvinAxion3,AshvinAxion1,AshvinAxion2,FanHOTI,FuKaneMele,FuKaneInversion,CohVDBAXI,VDBAxion}.  AXIs inherit their name from early work by Frank Wilczek~\cite{WilczekAxion} and Edward Witten~\cite{WittenDyon} on high-energy field theories that exhibit axion electrodynamics as a consequence of incorporating magnetic monopoles.  Signatures of axion electrodynamics were recently observed in terahertz spectroscopy probes of the 3D TI Bi$_2$Se$_3$ [\icsdweb{42545}, SG 166 ($R\bar{3}m$)] in an external magnetic field~\cite{WuAxionExp}.  Through a combination of magnetic SIs~\cite{EslamInversion,AshvinMagnetic,AshvinIndicators,MTQC,MTQCmaterials} and boundary-state calculations~\cite{TMDHOTI,WiederAxion,VDBAxion,KoreanAxion} of magnetically gapped, $\mathcal{I}$-symmetric TIs, it was revealed that AXIs are in fact magnetic HOTIs with intrinsic hinge states [Fig.~\ref{fig:HOTI}(b)].  Whereas the bulk topology of an $\mathcal{I}$-symmetric AXI may be inferred from the Fu-Kane parity formula~\cite{AshvinAxion2,AshvinAxion3,AndreiInversion}, all of the surface states and Wilson loops are generically gapped in an AXI with only bulk $\mathcal{I}$ symmetry, and the nontrivial bulk topology manifests as chiral spectral flow in the hinge and nested Wilson spectra~\cite{WiederAxion}.  Notably, there also exist AXI phases protected by bulk symmetries other than $\mathcal{I}$ -- such the product of twofold rotation and $\mathcal{T}$ -- in which the bulk topology cannot be diagnosed from SIs, and is only indicated by (nested) Wilson loop winding~\cite{WiederAxion,KoreanAxion,NicoDavidAXI2}.  AXI phases have been proposed in Sm-doped Bi$_2$Se$_3$~\cite{AxionZhida} and MnBi$_2$Te$_4$ [\icsdweb{425966}, SG 166 ($R\bar{3}m$)]~\cite{OtherAxion1,OtherAxion2}, and experimental signatures of AXIs have been observed in transport and ARPES probes of Cr- and V-doped (Bi,Sb)$_2$Te$_3$ heterostructures~\cite{OtherAxion3,OtherAxion4} and in MnBi$_2$Te$_4$~\cite{AxionExp1,AxionExp2}.

Lastly, by using TQC to re-express the original quadrupole insulator model as a magnetic obstructed atomic limit formed from $s$-$d$-orbital hybridization, researchers discovered the existence of solid-state higher-order TSM (HOTSM) phases with flat-band-like hinge states~\cite{HingeSM}.  In Dirac HOTSMs, there may or may not also exist topologically trivial Fermi-arc surface states; nevertheless, Dirac HOTSMs universally host intrinsic hinge states as a topological consequence of their bulk Dirac points~\cite{HingeSM,TaylorToy}.  Dirac HOTSM phases have since been predicted to occur in nearly all previously identified band-inversion Dirac semimetals~\cite{HingeSM}, and signatures of hinge states were recently observed in supercurrent oscillation experiments on the established Dirac semimetal Cd$_3$As$_2$ [\icsdweb{609930}, SG 137 ($P4_{2}/nmc$)]~\cite{HingeSMExp}.  Most recently, researchers have also predicted the existence of nodal-line~\cite{TMDHOTI,YoungkukMonopole}, Weyl~\cite{TaylorHigherOrderWeyl}, and sixfold-point~\cite{AndreiElectride} HOTSMs.  In HOTSM materials like Cd$_3$As$_2$, the bulk nodal degeneracies are equivalent to the quantum critical points between 2D insulators that differ by a quantized nested Berry phase, implying that the flat-band-like hinge states originate from higher-order-topological descent relations~\cite{HingeSM,TaylorToy,TMDHOTI,TaylorHigherOrderWeyl}.

\vspace{0.1in}
\section{The Topological Materials Database}
\label{sec:materials}

TQC~\cite{QuantumChemistry,JenFragile1,Bandrep1,Bandrep2,Bandrep3,BarryFragile} and related methods~\cite{SlagerSymmetry,AshvinIndicators,ChenTCI,AshvinTCI} have changed the mindset of the topological materials community, shifting the focus in materials discovery from human intuition to algorithmic materials prediction.  In particular, TQC has allowed the methods first introduced in~\cite{manes,SteveDirac,DDP,NewFermions,WiederLayers,WPVZprl} for predicting enforced semimetals and in~\cite{FuKaneInversion,ChenBernevigTCI} for predicting TIs and TCIs to be implemented for high-throughput topological materials discovery.  Exploiting the complete enumeration of symmetry-indicated TIs, TCIs, TSMs, HOTIs, HOTSMs, and atomic limits (elementary band coreps) developed over the past decade, researchers over the past two years have successfully performed the first large-scale searches for topological materials.  Utilizing the resources of the~\href{https://icsd.products.fiz-karlsruhe.de/}{ICSD}~\cite{ICSD1,ICSD2} and the~\href{https://materialsproject.org/}{Materials Project}~\cite{MatProject}, researchers applied the theories of TQC~\cite{QuantumChemistry,JenFragile1,Bandrep1,Bandrep2,Bandrep3,BarryFragile} and SIs~\cite{SlagerSymmetry,AshvinIndicators,ChenTCI,AshvinTCI} to the automated output of thousands of first-principles investigations of nonmagnetic chemical compounds, resulting in massive databases of stoichiometric topological materials~\cite{AndreiMaterials,ChenMaterials,AshvinMaterials1,AshvinMaterials2}.  Specifically, density functional theory~\cite{HohenbergKohn,KohnSham} calculations were performed on $\sim$25,000~\cite{AndreiMaterials}, $\sim$40,000~\cite{ChenMaterials}, and $\sim$20,000~\cite{AshvinMaterials1} of the~\TQCDTotICSDs~material structures entered in the ICSD after applying filters to remove alloys, entries with large numbers of atoms per unit cell, and materials containing heavy elements with $f$ electrons at $E_{F}$.  Around 8,000 ICSD entries were identified as stable TIs, TCIs, and TSMs at $E_{F}$ (intrinsic electronic filling).  Most recently, the first exhaustive high-throughput topological analysis of all of the stoichiometric materials in the ICSD was completed~\cite{AndreiMaterials2}.  In this section, we review the methodology and results of this analysis~\cite{AndreiMaterials2}, which are freely accessible on the~\webTQC~(\webNoICSD).

\begin{figure*}[!t]
\centering
\includegraphics[width=\textwidth]{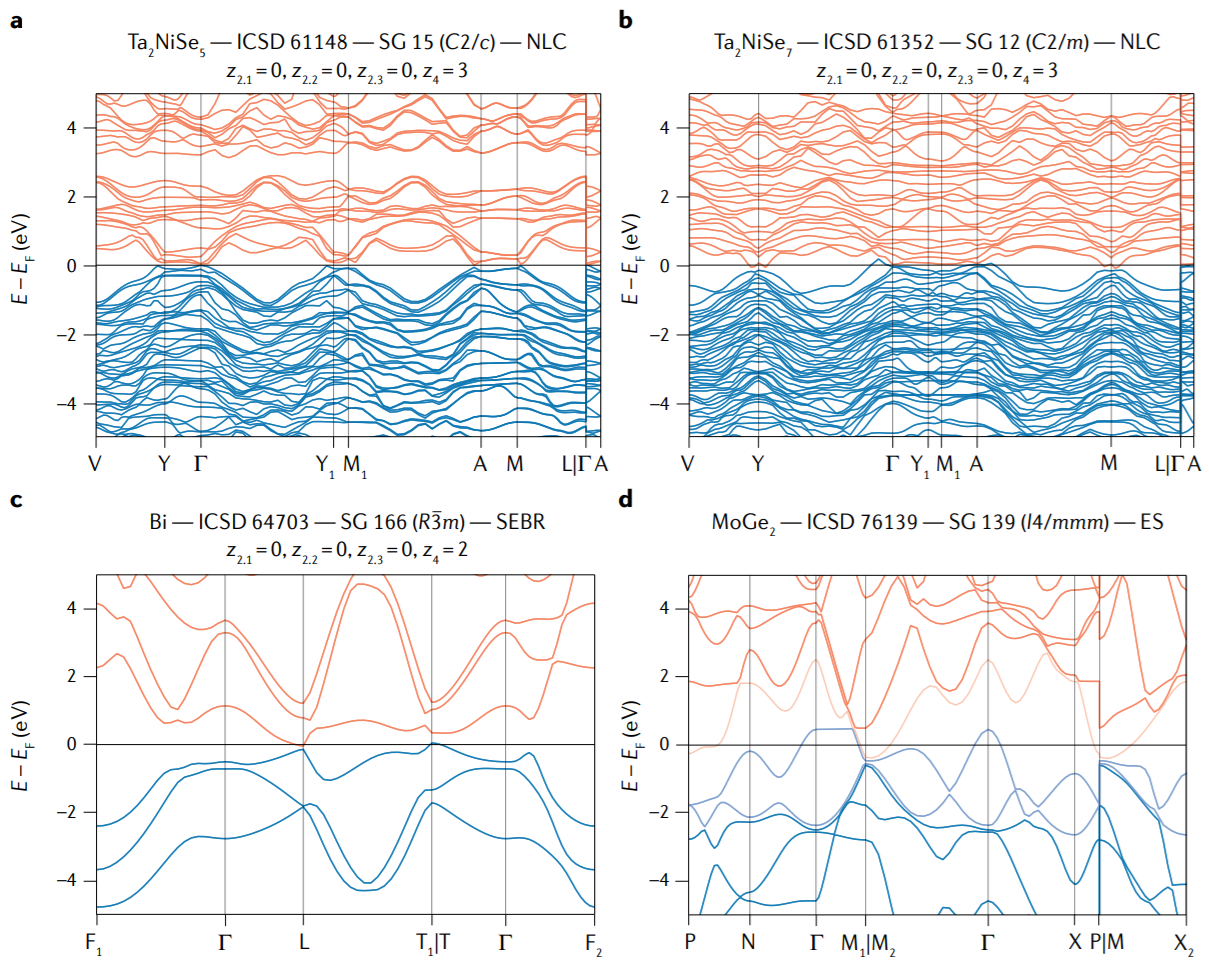}
\caption{Representative materials with novel topological properties identified in the~\webTQC.  For each material in this figure, we have provided the Inorganic Crystal Structure Database (ICSD) accession code and the number and standard-setting symbol of the crystallographic space group (SG) of the material~\cite{BigBook}.  For materials that are topological insulators (TIs) or topological crystalline insulators (TCIs) at intrinsic electronic filling, we additionally provide the cumulative symmetry-based indicators of the gap at the Fermi energy ($E_{F}$) in the notation established in~\cite{ChenTCI}.  (a) The layered transition-metal chalcogenide Ta$_2$NiSe$_5$ [\icsdweb{61148}, SG 15 ($C2/c$)] was measured in transport experiments to be an excitonic insulator~\cite{TaNiSeExciton1,TaNiSeExciton2}; in~\cite{AndreiMaterials2}, the narrow-gap semiconducting state of Ta$_2$NiSe$_5$ was revealed to be a $Z_{4}=3$ 3D TI.  Because its occupied bands cannot be expressed as a linear combination of disconnected branches of elementary band corepresentations (coreps), Ta$_2$NiSe$_5$ is classified as ``not equal to a linear combination'' (NLC).  (b) The closely-related quasi-1D transition-metal chalcogenide Ta$_2$NiSe$_7$ [\icsdweb{61352}, SG 12 ($C2/m$)], which conversely exhibits a charge-density-wave instability~\cite{Ta2NiSe7OriginalCDW}, was found in~\cite{AndreiMaterials2} to also be a $Z_{4}=3$, NLC-classified 3D TI in its normal state.  (c) Bi [\icsdweb{64703}, SG 166 ($R\bar{3}m$)] was previously recognized to be a $Z_{4}=2$ helical higher-order TCI~\cite{HOTIBismuth}; in~\cite{AndreiMaterials2}, it was discovered that the bands below $E_{F}$ in rhombohedral bismuth also exhibit stable topology.  Because the occupied bands of bismuth can be expressed as a linear combination of disconnected pieces of elementary band coreps, the bulk topology of bismuth originates from a split elementary band corep (SEBR).  (d) MoGe$_2$ [\icsdweb{76139}, SG 139 ($I4/mmm$)] was identified in~\cite{AndreiMaterials2} as a higher-order topological enforced semimetal (ES) with fourfold Dirac points near $E_{F}$ in which the two conduction and four valence bands closest to $E_{F}$ (doubly-degenerate light red and light blue bands) are as a set fragile topological.}
\label{fig:materials}
\end{figure*}

When the analysis in~\cite{AndreiMaterials2} was performed, the ICSD contained~\TQCDTotICSDs~entries, including multiple entries for the same material with varying structural parameters obtained in different experimental works.  First, the researchers excluded materials that were not close to the stoichiometric limit and entries with incorrect structure files (for example, files missing atoms present in the chemical formula), and then rounded nearly-stoichiometric materials to have integer-valued atomic occupancies.  This process reduced the initial~\TQCDTotICSDs~ICSD entries down to~\TQCDstoichiometric~stoichiometric entries with processable structure files.  The researchers then performed high-throughput density functional theory using the Vienna \emph{ab initio} Simulation Package (VASP)~\cite{VASP1,VASP2} on these~\TQCDstoichiometric~ICSD entries, both with and without incorporating the effects of SOC.  The high-throughput calculations resulted in convergent electronic structures for~\TQCDBNbrICSDs~ICSD entries in the presence of SOC.  To analyze the resulting electronic structures with TQC, the researchers applied the~\vasptotrace~tool~\cite{Vasp2TraceRef} that was previously implemented for the analysis in~\cite{AndreiMaterials}, and then wrote a new implementation of the~\webchecktopmat~tool on the~\webBCSshort.  The~\vasptotrace~program~\cite{Vasp2TraceRef} specifically outputs the small coreps that correspond to the Bloch states in each band at each ${\bf k}$ point, and~\webchecktopmat~determines whether an energy gap is permitted at a given electronic filling, and determines the stable and fragile SIs of all energetically isolated groups of bands.  After using~\vasptotrace~\cite{Vasp2TraceRef} and~\webchecktopmat~to analyze the~\TQCDBNbrICSDs~ICSD entries with convergent band-structure calculations, it was determined that the data contained~\TQCDBNbrUniqueMaterials~unique materials, defined such that two ICSD entries with the same chemical formulas, crystal structures, and bulk topology at $E_{F}$ were associated to the same unique material.  The computational calculations performed to generate the~\webTQC~required~\TQCDBVASPTotalCPUTimeAllMillionHours~million CPU hours.

For each ICSD entry in the~\webTQC, the topology is characterized using TQC from two perspectives:  cumulative band topology and energetically isolated topological bands.  In the first approach, given the number of electrons in each unit cell, the topology at $E_{F}$ is calculated using TQC and SIs, and is then sorted into one of the five broad categories defined in~\cite{AndreiMaterials}.  First, if the occupied bands of the material can be expressed as a linear combination of elementary band coreps, the material is classified as LCEBR.  LCEBR materials, in general, are either trivial or fragile topological, or host non-symmetry-indicated stable topology; in~\cite{AndreiMaterials2}, the authors did not observe any materials for which the entire valence manifold was fragile topological.  Next, if the symmetry eigenvalues of the occupied bands imply the existence of a TSM phase, the material is labeled as an enforced semimetal with Fermi degeneracy (ESFD) if the bulk is a high-symmetry-point TSM, and is labeled as an enforced semimetal (ES) if the bulk is a high-symmetry-line or plane TSM.  Lastly, if a gap is permitted at $E_{F}$ and the bulk exhibits nontrivial stable SIs, the occupied bands are either classified as a split elementary band corep (SEBR), or as ``not equal to a linear combination'' (NLC), depending on whether the occupied bands can or cannot be expressed as linear combinations of pieces (disconnected branches) of elementary band coreps.  Of the~\TQCDBNbrUniqueMaterials~unique materials,~\TQCDBNbrMaterialstrivial~(\TQCDBNbrMaterialsLCEBRPercent)~were LCEBR,~\TQCDBNbrMaterialsSM~(\TQCDBNbrMaterialsSMPercent)~were enforced TSMs at $E_{F}$, and~\TQCDBNbrMaterialsTI~(\TQCDBNbrMaterialsTIPercent)~were TIs and TCIs with nontrivial stable SIs.  These results surprisingly implied that~\TQCDBNbrTopologicalMaterialsPercent~of the known stoichiometric 3D materials are symmetry-indicated TIs, TCIs, or TSMs at intrinsic filling ($E_{F}$).  Notably, this is considerably higher than the percentage of topological stoichiometric 2D materials, which has been computed to lie within the range of a few to 20\%~\cite{Marzari2DExfoliateDB,Olsen2DTISearch,Marzari2DTIAbundantDB,Ashvin2DMaterials}.

Unlike in previous high-throughput topological analyses~\cite{AndreiMaterials,ChenMaterials,AshvinMaterials1,AshvinMaterials2}, the version of~\webchecktopmat~implemented for~\cite{AndreiMaterials2} allowed researchers to analyze the symmetry-indicated stable and fragile topology of energetically isolated bands away from $E_{F}$.  This enabled the authors of~\cite{AndreiMaterials2} to employ a second approach towards topological materials characterization.  For each ICSD entry in the~\href{https://www.topologicalquantumchemistry.com/}{Topological Materials Database}, the researchers additionally calculated the stable and fragile topology of each set of isolated bands in the energy spectrum as defined by band connectivity through TQC~\cite{QuantumChemistry,JenFragile1,Bandrep1,Bandrep2,Bandrep3}.  Discarding bands from core-shell atomic orbitals, it was discovered that among all of the isolated bands in all of the~\TQCDBNbrUniqueMaterials~analyzed unique materials,~\TQCDBPercentFragileBandSetsUniqueMaterials~of bands are fragile topological,~\TQCDBPercentStrongBandSetsUniqueMaterials~of bands -- nearly 2/3 -- are stable topological, and an overwhelming~\TQCDBNbrTopoBandPercent~of known stoichiometric materials contain at least one stable or fragile topological band.  These numbers are all the more shocking when one considers that many of the unique materials in the~\href{https://www.topologicalquantumchemistry.com/}{Topological Materials Database} respect the symmetries of one of the 117 noncentrosymmetric, nonmagnetic SGs in which stable topology cannot be determined by SIs~\cite{ChenTCI,AshvinTCI}.  Indeed, the analysis in~\cite{AndreiMaterials2} has revealed that band topology is one of the most fundamental and generic properties of solid-state materials.

Each of the three research groups that performed the high-throughput searches for topological materials~\cite{AndreiMaterials,ChenMaterials,AshvinMaterials1,AshvinMaterials2,AndreiMaterials2} implemented a website for users to access the resulting data.  In addition to the aforementioned~\webTQC~implemented for~\cite{AndreiMaterials,AndreiMaterials2}, high-throughput first-principles topological material calculations can be accessed on the~\href{http://materiae.iphy.ac.cn/}{Catalogue of Topological Materials} (\url{http://materiae.iphy.ac.cn/}) created for~\cite{ChenMaterials} and the~\href{https://ccmp.nju.edu.cn/}{Topological Materials Arsenal} (\url{https://ccmp.nju.edu.cn/}) created for~\cite{AshvinMaterials1,AshvinMaterials2}.  All three websites allow users to search for materials by SG and chemical formula, with further functionality and more complete topological characterizations being dependent on the particular website.  The~\webTQC~and the~\href{http://materiae.iphy.ac.cn/}{Catalogue of Topological Materials} in particular allow users to perform more advanced queries, including searches restricted to materials with specific gap sizes and stable SIs.  The~\webTQC~additionally provides users with the stable and fragile topological classification of energetically isolated bands away from $E_{F}$.  In Fig.~\ref{fig:materials}, we show examples of materials in the~\webTQC~identified in~\cite{AndreiMaterials2} as hosting novel topological properties.

\section{Conclusion and Future Directions}
\label{sec:conclusion}

In this Review, we have followed the discovery of solid-state topological materials from the first theoretical proposals of $\mathcal{T}$-symmetric TIs to the recent identification of tens of thousands of candidate TIs, TCIs, and TSMs through high-throughput searches.  Despite the recent completion of nonmagnetic solid-state group theory~\cite{QuantumChemistry,JenFragile1,Bandrep1,Bandrep2,Bandrep3,AshvinIndicators,ChenTCI,AshvinTCI}, and the complete analysis of the stoichiometric materials in the ICSD performed in~\cite{AndreiMaterials2}, there remain many unexplored directions for topological materials discovery.  First, the methods used to identify topological materials in~\cite{AndreiMaterials,AndreiMaterials2,ChenMaterials,AshvinMaterials1,AshvinMaterials2,MTQCmaterials} can only be applied to materials with nontrivial stable and fragile SIs, which are largely centrosymmetric.  However, the ICSD~\cite{ICSD1,ICSD2} contains thousands of noncentrosymmetric compounds.  Because it was discovered~\cite{AndreiMaterials2} that~\TQCDBNbrTopoBandPercent~of the stoichiometric materials in the ICSD contain topological bands when only SIs are taken into account, there likely exist thousands of yet-unidentified noncentrosymmetric TIs, TCIs, and TSMs.  Furthermore, whereas the complete stable SIs of magnetic and nonmagnetic TIs, TCIs, and TSMs have been computed~\cite{AshvinIndicators,AshvinMagnetic,ZhidaSemimetals,AshvinTCI,ChenTCI,MTQC,MTQCmaterials}, the magnetic fragile SIs remain an outstanding problem.  One route towards the discovery of topological materials beyond SIs involves developing a high-throughput protocol based on carefully selected Wilson loop calculations; several promising early efforts in this direction have been completed this past year~\cite{S4Weyl1,S4Weyl2}.  Additionally, despite recent early efforts~\cite{HingeSM,TMDHOTI,WiederAxion,HourglassInsulator,Cohomological,DiracInsulator}, there remain several 3D TCI phases without SIs that have been identified through layer constructions and surface state calculations~\cite{MTQC,ChenTCI}, but do not yet have accompanying bulk [(nested) Wilson loop] topological invariants.

Second, TQC was extended this past year to crystals with commensurate magnetism to construct the theory of magnetic TQC (MTQC)~\cite{MTQC}.  The formulation of MTQC required the intermediate computation of the magnetic small coreps and elementary band coreps, completing the century-old problem of enumerating irreducible coreps in solid-state group theory~\cite{sohncke1879book,fedorov1891symmetry,ShubnikovBook}.  Through MTQC, the complete SI groups and formulas of spinful band topology in all 1,651 magnetic and nonmagnetic SGs were determined~\cite{MTQC}, subsuming an earlier calculation of the magnetic SI groups~\cite{AshvinMagnetic}, and leading to the discovery of several novel variants of magnetic HOTIs.  MTQC was most recently employed to perform a high-throughput search for magnetic topological materials~\cite{MTQCmaterials}, resulting in the identification of over 100 magnetic TIs, TCIs, and TSMs among the $\sim$500 magnetic materials on the~\webBCSshort~with experimentally measured magnetic SGs~\cite{BCSMag1,BCSMag2,BCSMag3,BCSMag4}.  Structurally chiral magnetic TSM phases were also recently proposed in chiral crystals with commensurate magnetic ordering, including Mn$_3$IrSi [\icsdweb{151748}, SG 198 ($P2_{1}3$)]~\cite{MagneticNewFermion}.  However, despite these results, and the recent experimental identification of novel magnetic TCIs~\cite{AxionExp1,AxionExp2,OtherAxion3,OtherAxion4,AxionZahid1} and TSMs~\cite{MagneticWeylZahid,MagneticWeylYulin,MagneticWeylHaim}, the number of materials whose commensurate magnetic structures have been measured through neutron diffraction and matched with magnetic SGs remains several orders of magnitude smaller than the number of nonmagnetic materials with measured crystal structures and SGs.  It is our hope that the recent successes of MTQC and magnetic SIs will encourage experimentalists and computational material scientists to analyze larger numbers of magnetic structures using the magnetic SG classification of magnetic crystalline solids~\cite{BigBook}.  This symmetry analysis would facilitate future theoretical and experimental investigations of materials with linked magnetic and topological order, both at the mean-field level of MTQC~\cite{MTQC} and beyond.

As a third future direction, many of the compounds in the ICSD feature unreported structural phases or chemical complexities beyond the simple descriptions provided in their structure files.  A hybrid methodology combining the intuition of trained chemists with high-throughput calculations of topological electronic structures~\cite{AndreiMaterials,AndreiMaterials2,ChenMaterials,AshvinMaterials1,AshvinMaterials2} could facilitate the development of advanced alloyed or heterostructure topological materials.  The engineering of topological heterostructure and few-layer devices in particular has emerged as a promising direction for future studies.  For example, researchers have experimentally demonstrated an electronically switchable photocurrent response in the 2D TI WTe$_2$~\cite{SuyangPabloCPGEWTe2} and cuprate-like superconductivity in twisted bilayer graphene~\cite{BLGSC1,BLGSC2,BLGSC3}.  Most intriguingly, recent theoretical~\cite{ZhidaBLG,AshvinBLG1,AshvinBLG2,KoreanFragile,WiederAxion,HigherOrderTBLG,TBLGSuperfluidWeight,HofstadterTheory1,HofstadterTheory2} and experimental~\cite{TBLGnewExp1,TBLGnewExp2,TBLGnewExp3,TBLGnewExp4,TBLGnewExp5,TBLGnewExp6,TBLGnewExp7,TBLGnewExp8} studies have revealed a relationship between the superconductivity and other interacting phases in twisted bilayer graphene and both fragile and higher-order topology.  Further investigations into few-layer topological device engineering may therefore provide insight into some of the largest outstanding problems in solid-state physics, such as the mechanism of superconductivity in the high-temperature cuprate superconductors~\cite{PhilCuprateGlue}.

Lastly, though this Review is largely focused on noninteracting topological phases with crystal symmetry, an emerging future direction in topological materials involves the experimental manipulation of charge-density-wave (CDW) and spin-density-wave phases in solid-state materials.  For example, recent theoretical and experimental investigations have provided evidence that structurally chiral, quasi-1D (TaSe$_4$)$_2$I crystals [\icsdweb{35190}, SG 97 ($I422$)] are Weyl semimetals that become gapped by an incommensurate CDW~\cite{CDWWeyl}.  In earlier works, Weyl-CDW insulators were proposed to be correlated AXIs~\cite{ShouchengCDW,TitusCDW,TaylorCDW}, though recent investigations have revealed that the axionic response of Weyl-CDWs is captured by mean-field theory~\cite{WiederBradlynCDW,JiabinWeylCDW,KuansenWeylCDW}.  The prediction~\cite{CDWWeyl} of a Weyl-CDW phase in (TaSe$_4$)$_2$I was recently confirmed through ARPES measurements of high-temperature Weyl points~\cite{OtherCDWARPES} and transport experiments demonstrating axionic response effects in the low-temperature CDW phase~\cite{AxionCDWExperiment}.  As discussed in earlier in this Review, AXIs have recently been recognized to be HOTIs; additionally, recent theoretical investigations~\cite{GilCDWHOTI} of the CDW phase in 1T-TaS$_2$ [\icsdweb{52115}, SG 164 ($P\bar{3}m1$)]~\cite{TMDCDWAbbamonte} have demonstrated the presence of higher-order topological corner states.  $\mathcal{T}$-symmetric TSM-CDWs are therefore likely to provide a promising platform for elucidating and exploring novel response effects in helical TCIs and HOTIs, phases of matter that owe their existence to the interplay of symmetry and topology in solid-state materials.

\vspace{0.1in}
\centerline{\bf Acknowledgments}
\vspace{0.1in}
This review is dedicated to Prof. A. A. Soluyanov, who passed away during its preparation.  $^\dag$Corresponding author: \url{bwieder@mit.edu} (B. J. W.), \url{bernevig@princeton.edu} (B. A. B.).  $^\ddag$Primary address.  B. J. W., N. R., and B. A. B. were supported by the Department of Energy Grant No. DE-SC0016239, NSF EAGER Grant No. DMR 1643312, NSF-MRSEC Grant No. DMR-142051, Simons Investigator Grant No. 404513, ONR Grant No. N00014-20-1-2303, the Packard Foundation, the Schmidt Fund for Innovative Research, the BSF Israel US Foundation Grant No. 2018226, the Gordon and Betty Moore Foundation through Grant No. GBMF8685 towards the Princeton theory program, and a Guggenheim Fellowship from the John Simon Guggenheim Memorial Foundation.  B. B. acknowledges the support of the Alfred P. Sloan Foundation and the National Science Foundation Grant No. DMR-1945058.  J. C. acknowledges support from the National Science Foundation Grant No. DMR 1942447 and the Flatiron Institute, a division of the Simons Foundation.  Z. W. was supported by the National Natural Science Foundation of China [Grant No. 11974395], the Strategic Priority Research Program of the Chinese Academy of Sciences (CAS) [Grant No. XDB33000000], and the Center for Materials Genome.  M. G. V. acknowledges support from DFG INCIEN2019-000356 from Gipuzkoako Foru Aldundia and the Spanish Ministerio de Ciencia e Innovacion (Grant No. PID2019-109905GB-C21).  L. E. was supported by the Government of the Basque Country (Project IT1301-19) and the Spanish Ministry of Science and Innovation (PID2019-106644GB-I00).  C. F. was supported by the ERC Advanced Grant No. 291472 ‘Idea Heusler’ and ERC Advanced Grant No. 742068 ‘TOPMAT’.  T. N. acknowledges support from the European Union's Horizon 2020 Research and Innovation Program (ERC-StG-Neupert-757867-PARATOP).  A. S. and T. N. additionally acknowledge support from the Swiss National Science Foundation under Grant No. PP00P2\_176877.  L. E., N. R., and B. A. B. acknowledge additional support through the ERC Advanced Grant Superflat, and B. A. B. received additional support from the European Union's Horizon 2020 Research and Innovation Program (Grant No. 101020833) and the Max Planck Society.

\vspace{0.05in}
\centerline{\bf Author Contributions}
\vspace{0.05in}

This Review was written and edited by B. J. W.  All authors contributed to researching data for this article and to the preparation of the manuscript.

\vspace{0.05in}
\centerline{\bf Competing Interests}
\vspace{0.05in}

The authors declare no competing interests.

\bibliography{TM_NM}

\end{document}